%% file: main.tex
\documentclass[journal,twoside,web]{ieeecolor}

\usepackage{algorithm}
\usepackage{amsmath,amssymb,amsfonts}
\usepackage{bbold}
\usepackage{cite}
\usepackage[mathscr]{euscript}
\usepackage{generic}
\usepackage{graphicx}
\usepackage{textcomp}
\usepackage{booktabs}
\usepackage{tabularx}
\usepackage{adjustbox}
\usepackage{algpseudocode}
\usepackage{placeins}
\usepackage{enumerate}
\usepackage{capt-of}
\usepackage{hyperref}

\def\BibTeX{{\rm B\kern-.05em{\sc i\kern-.025em b}\kern-.08em
    T\kern-.1667em\lower.7ex\hbox{E}\kern-.125emX}}

\makeatletter
\def\@opargbegintheorem#1#2#3{%
  \@IEEEtmpitemindent\itemindent\relax
  \topsep 0pt
  \rmfamily
  \trivlist
  \item[]\textit{\indent #1\ #2\ (#3):}%
  \itemindent\@IEEEtmpitemindent\relax
}
\makeatother
\renewcommand{\eqref}[1]{(\ref{#1})}  

\newtheorem{definition}{Definition}
\newtheorem{remark}{Remark}
\newtheorem{theorem}{Theorem}
\newtheorem{lemma}{Lemma}

\newtheorem{fact}{Fact}

\newcommand{\card}{\mathrm{card}}
\newcommand{\V}{\mathcal{V}}
\newcommand{\diag}{\mathrm{diag}}
\newcommand{\PA}{\mathrm{PA}}
\newcommand{\A}{\mathcal{A}}
\newcommand{\F}{\mathcal{F}}
\newcommand{\bfU}{\mathbf{U}}

\newcommand{\OOmega}{\mathbf{\Omega}}
\newcommand{\C}{\mathcal{C}}
\newcommand{\TTheta}{\mathbf{\Theta}}
\newcommand{\bfS}{\mathbf{S}}
\newcommand{\Aid}{\mathrm{Aid}}
\newcommand{\sgn}{\mathrm{sgn}}
\newcommand{\zcz}[1]{{#1}}
\hypersetup{hidelinks}

\title{Power Allocation Games on Signed Networks: Nash Equilibria and Coevolutionary Dynamics}
\author{Chuanzhe Zhang$^{1}$, Yuke Li$^{2}$, and Wenjun Mei$^{1}$
\thanks{This work was supported in part by the National Natural Science Foundation of China under Grants 72201008, 72531001, and 72131001, the National Key R\&D Program of China under Grant 2022ZD0116401 and 2022ZD0116400.}
\thanks{An earlier conference version of this manuscript appeared in the Proceedings of the 62nd \emph{IEEE Conference on Decision and Control}. The present journal version contains substantial new material, including (i) complete proofs and additional analyses for the static game model, (ii) a new section on power–strategy coevolutionary dynamics, and (iii) expanded numerical simulations and empirical studies supporting the proposed dynamic model.}
\thanks{$^{1}$Chuanzhe Zhang and Wenjun Mei are with the Department of Control and Systems Engineering, Peking University, Beijing, China. $^{2}$Yuke Li is with the Norman Paterson School of International Affairs, Carleton University, Ottawa, Canada.
}
\thanks{The corresponding author is Wenjun Mei (mei@pku.edu.cn).}
}

\begin{document}

\maketitle
\thispagestyle{empty}

\begin{abstract}
Understanding how strategic interactions and power distributions coevolve in international relations is central to explaining conflict, cooperation, and long-term inequality. We study this problem using a power-allocation game on signed networks. Departing from models that restrict strategy updates to Pareto improvements, we propose a generalized formulation in which countries prioritize self-survival and strategically trade off between supporting allies and weakening adversaries. This relaxation allows countries to sacrifice certain allies to achieve higher overall payoffs. For the resulting static game, we establish the existence of pure-strategy Nash equilibria and characterize their properties in extreme cases, including fully antagonistic networks and the presence of a dominant power. We further introduce a power–strategy coevolutionary dynamic and prove its almost-sure convergence to equilibria corresponding to the static game. The proposed models are validated using empirical data and numerical simulations. Historical data from the Correlates of War and national capability datasets show that survival likelihood predicts countries’ safety outcomes and subsequent economic growth with relatively high accuracy. Simulations further indicate that, under fixed conflict intensity, more structurally balanced signed networks yield higher average power and lower inequality at steady states.
\end{abstract}


\section{Introduction}
\subsection{Background and motivation}
In recent years, international relations research has experienced a methodological shift from qualitative analysis to empirical and quantitative approaches~\cite{deutsch1964multipolar,cranmer2015kantian}. Network games provide an analytical framework for capturing the strategic interactions underlying real-world international relations. Despite this potential, relatively few studies model international relations explicitly as games and dynamics on complex networks.

To better capture the complexity of strategic decision making in international relations, this paper proposes and analyzes a generalized game model with a key conceptual advancement: countries are allowed to act in a more strategic and realistic manner by explicitly weighing gains against losses when sacrificing allies or reconciling with enemies. Based on this static game, we further introduce a power–strategy coevolutionary dynamic, enabling a systematic investigation of how signed network structure shapes global prosperity and inequality.

\subsection{Literature review}
As advocated by leading political scientists such as K. Deutsch~\cite{deutsch1963nerves,deutsch1968toward,deutsch1975some}, the use of formal modeling has become an important methodological trend in the study of international relations~\cite{PDH-MDW:04,AB-SC:11,bryen2012application}. However, early game-theoretic studies in international relations typically focused on games with only two or three players or strategies~\cite{kydd2003side,correa2001game}. Other works relied primarily on network analysis~\cite{goddard2009brokering,hafner2009network}, but often lacked a formal game-theoretic structure~\cite{decanio2013game,jervis1988realism}. This gap highlights the need for more general modeling frameworks that can capture strategic behavior in complex international-relation networks.

Many core problems in international relations can be formulated as network games, where countries strategically allocate limited resources or power and their utilities depend on both their own and others' allocations~\cite{bjorner1991chip,bjorner1992chip,chaudhuri2008network,kalai1982generalized,myerson1977graphs,dubey1984totally,pacheco2006coevolution}. Although network resource allocation has been extensively studied in domains such as transportation and wireless communication~\cite{bredin2000game,etesami2014optimal,johari2004efficiency,cominetti2006network}, most existing models are not directly applicable to international relations.

The model proposed in this paper is most closely related to a series of works by Li and Morse~\cite{li2017countries,li2018power,li2022games}, which formulate international relations as power-allocation games on signed networks, with positive and negative edges representing friendly and antagonistic relations. Countries allocate power among self-protection, supporting allies, and attacking enemies. A country’s safety state: safe, precarious, or in danger, depends on its own allocation as well as support and attacks from others. Li and Morse~\cite{li2022games} establish the existence of Nash equilibria and analyze the coevolution of power-allocation strategies and signed network structures. However, their framework restricts strategy updates to Pareto improvements in neighbors’ safety states, thereby precluding strategic trade-offs that sacrifice some allies for greater overall gains and limiting its ability to capture real-world decision making.

\subsection{State of contributions}
This paper makes three main contributions.

First, we generalize the power-allocation game by allowing countries to prioritize self-survival and then strategically trade off between supporting allies and attacking enemies. Thus, a country may sacrifice some allies when doing so yields a higher overall payoff. We also extend the static game to power–strategy coevolutionary dynamics, in which countries’ power levels evolve endogenously according to their safety states. This dynamic framework enables systematic analysis of how the signed network structure of international relations shapes long-term prosperity and inequality.

Second, we conduct a comprehensive theoretical analysis of both the static game and the coevolutionary dynamics. We prove the existence of pure-strategy Nash equilibria for the static power-allocation game and analytically characterize their structural properties in several extreme cases, including fully antagonistic networks and the presence/absence of a dominant power. For the dynamic model, we establish almost-sure convergence to an equilibrium, which is also a Nash equilibrium of the static game. 

Third, we validate the proposed models using empirical data and numerical simulations. For the static game, we introduce the notion of survival likelihood and show that, when calibrated with Correlates of War and the Composite Index of National Capability (CINC) data during World War II, it predicts countries’ safety outcomes with annual accuracy exceeding 70\%. For the dynamic model, we demonstrate that survival likelihood reliably predicts subsequent GDP growth using data from 1955–2002, with average accuracy above 80\%. Finally, simulations of the coevolutionary dynamics reveal a clear structural pattern: given fixed edge density and conflict intensity, more structurally balanced signed networks lead to higher average national power and lower inequality (as measured by the Gini coefficient) at steady states.

\subsection{Organization}
The rest of this paper is organized as follows: Section II introduces and studies the static power-allocation game, including model setup, existence of Nash equilibria, and analysis of some special scenarios. Section III extends the static game model to coevolutionary power-strategy dynamics and proves almost-sure convergence. Section IV presents numerical and empirical results. Section V concludes the paper.

\section{Static model: power-allocation games on signed networks}
\subsection{Basic notations and definitions}\label{subsection:notations}
Let $\subseteq$ and $\subset$ be the symbols for subset and proper subset respectively. Denote by $\mathbb{N}_{+}$ the set of positive integers and let $\mathbb{N}=\mathbb{N}_{+}\cup \{0\}$. Denote by $\card(\cdot)$ the number of elements in a set. Let $\mathbb{1}_{\{\cdot\}}$ be the indicator function, i.e., $\mathbb{1}_{\{C\}}=1$ if the condition $C$ holds, and $\mathbb{1}_{\{C\}}=0$ otherwise. Let $\mathbf{1}_n$ be the column vector of ones, i.e., $\mathbf{1}_n=(1,1,\dots,1)^{\top}$. Denote by $\diag(x)$ the diagonal matrix with the entries of the vector $x$ on its diagonal. Define the sign function $\sgn(x)$ as $\sgn(x)=1$ if $x>0$; $\sgn(x)=0$ if $x=0$; $\sgn(x)=-1$ if $x<0$.

\subsection{Model setup: static power-allocation game}\label{subsection:static-model-setup}
In this section, we formalize a power-allocation game on signed networks in the context of international relations. To distinguish it from the dynamic model introduced in the next section, we refer to the model studied here as the \emph{\textbf{s}tatic \textbf{p}ower-\textbf{a}llocation game} (SPA game).

\textit{International relations as a signed graph:} Consider $n$ countries indexed by $i\in\V=\{1,\dots,n\}$. Although international relations are dynamic in practice, we focus on the role of network structure in shaping Nash equilibria and power evolution and therefore assume fixed relations. For each country $i$, let $\F_i$ and $\A_i$ denote its sets of friendly and antagonistic countries (friends and enemies for short), respectively. We assume that $\F_i\cap\A_i=\varnothing$ for all $i\in\V$. All relations are bilateral and symmetric: $j\in\F_i$ if and only if $i\in\F_j$, and $j\in\A_i$ if and only if $i\in\A_j$. For convenience, each country is included in its own friend set, i.e., $i\in\F_i$ for all $i\in\V$. The resulting structure forms an undirected, unweighted signed graph, where nodes represent countries and positive (negative) edges represent friendly (antagonistic) relations. Throughout the paper, the terms ``graph'' and ``network'' are used interchangeably.

\textit{Power-allocation strategies: }For each country $i$, let its power be a fixed nonnegative integer $p_i$. Country $i$ allocates its power either to attack its enemies or to support its friends. For any $j\in\A_i$ (or $j\in\F_i$ respectively), let $x_{ij}\in \mathbb{N}$ denote the amount of power that country $i$ spends on attacking (or supporting respectively) $j$. 
The row vector $(x_{i1},\dots,x_{in})\in \mathbb{N}^{1\times n}$ is referred to as the strategy of country $i$. Denote by $X=(x_{ij})_{n\times n}$ the strategies of all countries and call it a \emph{strategy matrix}. By definition, 
\begin{align*}
X\in \TTheta = \Big{\{}X\in \mathbb{N}^{n\times n}\,\Big|\, & \text{For any }i,\ \sum_{j\in \A_i\cup \F_i} x_{ij}=p_i\\
&\text{and }x_{ij}= 0\text{ if }j\notin \A_i\cup \F_i\Big{\}}.
\end{align*}
The integer-valued formulation is motivated by the fact that many resources contributing to national power, such as personnel, military units, and equipment, are naturally indivisible. For inherently continuous resources, it can be interpreted as a discretization at a prescribed resolution. Thus, $p_i$ and $x_{ij}$ are modeled as integer-valued quantities throughout the paper.

\textit{States of countries:} Given a strategy matrix $X$, each country's safety state is defined as follows.
\begin{definition}[Safety states]\label{def:state}
For any $X\in\mathbb{N}^{n\times n}$ and $i\in\V$, define
\begin{align*}
    s_i(X)=\sum_{j\in \A_i} x_{ij} + \sum_{j\in \F_i} x_{ji} - \sum_{j\in \A_i} x_{ji}.
\end{align*}
Country $i$ is \emph{safe} if $s_i(X)>0$, \emph{precarious} if $s_i(X)=0$, and \emph{in danger} if $s_i(X)<0$.
\end{definition}
Definition~1 captures the direct effects of a
given power-allocation profile on countries' safety states. To put it simply, a country's safety state is determined by its offensive effort, support from friends, and attacks from enemies.

\textit{Preference axioms: }To make our model more general and inclusive, we do not specify countries' utility functions, but instead assume that they obey the following preference axiom.
\begin{definition}[Preference]\label{def:preference}
Given two strategy matrices $X$ and $Y$, country $i\in \{1,\dots, n\}$ prefers $X$ to $Y$, denoted by $X\ge_i Y$, if either of the following two conditions holds:
\begin{enumerate}
\item $s_i(X)\ge 0$ and $s_i(Y)<0$;
\item Either $s_i(X), s_i(Y)\ge 0$ or $s_i(X), s_i(Y)<0$. Moreover, 
\begin{align*}
&\quad\sum_{j\in \A_i}\mathbb{1}_{\{s_j(X)\leq 0\}}+\sum_{l\in \F_i} \mathbb{1}_{\{s_l(X)\ge 0\}} \\
& \ge \sum_{j\in \A_i} \mathbb{1}_{\{s_j(Y)\leq 0\}}+\sum_{l\in \F_i} \mathbb{1}_{\{s_l(Y)\ge 0\}}.
\end{align*}

\end{enumerate}
\end{definition}

Definition 2 captures a hierarchy of strategic objectives. A country first prioritizes its own survival and therefore prefers a strategy matrix under which it is not in danger to one under which it is in danger. If its own safety states under two strategy matrices belong to the same category, it prefers the matrix with a larger total number of friends that are not in danger and enemies that are not safe. Moreover, it is straightforward to verify that, for each country, the resulting preference relation is complete and transitive, meaning that any two strategy matrices are comparable and the ranking is consistent across comparisons.

Compared with the model in~\cite{li2022games}, the static game studied here introduces a key conceptual advance by allowing more flexible and realistic strategic behavior. In~\cite{li2022games}, countries revise their strategies only through Pareto improvements that preserve all neighbors’ benefits. In contrast, our model relaxes this restriction by permitting countries to sacrifice the safety of certain friends when doing so increases their overall payoff, which better reflects real-world decision making.

\begin{remark}
There exist various utility functions compatible with the axioms above, e.g., the utility $u_i(X) =  n\mathbb{1}_{\{s_i(X)\ge 0\}} + \sum_{j\in \F_i} \mathbb{1}_{\{s_j(X)\ge 0\}} + \sum_{j\in \A_i} \mathbb{1}_{\{s_j(X)\leq 0\}}$ for each $i\in \V$. 
\end{remark}

\textit{Nash equilibrium:} By Definition~\ref{def:preference}, a strategy matrix $X^*\in \TTheta$ is called a pure-strategy Nash equilibrium if, for any $i\in\V$, $X^*\ge_i X$ for any $X\in \TTheta$ that may differ from $X^*$ only in the $i$-th row. 

\subsection{Existence and computation of Nash equilibrium}
In this section, we establish one of the main results of this paper. That is, any game satisfying the preference axioms given by Definition~\ref{def:preference} admits at least one pure-strategy Nash equilibrium. The theorem is formally presented as follows.

\begin{theorem}[Existence of Nash Equilibrium]\label{thm:existence-NE}
Given $n$ countries with powers $p_1,\dots,p_n$ and their respective sets of friendly and antagonistic countries $\F_1,\A_1,\dots,\F_n,\A_n$, the static power-allocation game defined in Section~II-B admits at least one special pure-strategy Nash equilibrium. In this equilibrium, no country is in danger, and no power is allocated between any two distinct friendly countries.
\end{theorem}

The proof proceeds in three steps, outlined as follows:
\begin{enumerate}
\item Construct a subset of strategy matrices, termed the no-support-no-danger (NSND) matrices.
\item Define an iterative procedure starting from an NSND matrix and based on an operation called the \textit{preferable adjustment}.
\item Show that this iteration must terminate in finite time, and that the resulting strategy matrix constitutes a Nash equilibrium of the static power-allocation game.
\end{enumerate}

Following the above outline, several definitions and facts needed for the proof are introduced first. 
\begin{definition}[NSND strategy matrix]\label{def:NSND-strategy-matrix}
Consider the static power-allocation game defined in Section~II-B. A strategy matrix $X\in \TTheta$ is called a no-support-no-danger (NSND) strategy matrix, if it satisfies the following conditions: 1) $X=X^{\top}$; 2) $x_{ij}=0$ for any $i,\,j$ such that $j\notin \A_i\cup \{i\}$.
\end{definition}

The following fact can be easily inferred from the definition of countries’ states and the preference axioms.

\begin{fact}\label{fact:NSND-properties}
For any NSND strategy matrices $X$ and $Y$ and any $i\in\V$, the following properties hold:
\begin{enumerate}
    \item $s_{i}(X)=x_{ii}\ge 0$, which means that no country is in danger under any NSND strategy matrix.
    
    \item $X\ge_i Y$ if and only if 
    \begin{align*}
        \sum_{j\in \A_i} \mathbb{1}_{\{x_{jj}=0\}} & = \sum_{j\in \A_i}  \mathbb{1}_{\{s_j(X)=0\}} \\
        & \ge \sum_{j\in \A_i} \mathbb{1}_{\{s_j(Y)=0\}} = \sum_{j\in \A_i} \mathbb{1}_{\{y_{jj}=0\}}.
    \end{align*}
    Namely, $i$ prefers an NSND matrix $X$ to another NSND matrix $Y$ if $i$ has more precarious enemies under $X$.
\end{enumerate}
\end{fact}

\begin{proof}
For Statement 1), according to Definition~\ref{def:NSND-strategy-matrix}, $X$ is symmetric and $x_{ji}=0$ for every $j\in\F_i\setminus\{i\}$. Hence,
\begin{align*}
s_i(X)
&=
\sum_{j\in\A_i}x_{ij}
+\sum_{j\in\F_i}x_{ji}
-\sum_{j\in\A_i}x_{ji} \\
&=
\sum_{j\in\A_i}x_{ij}
+x_{ii}
-\sum_{j\in\A_i}x_{ji}
=x_{ii}\geq 0.
\end{align*}
Therefore, every country is either safe or precarious under $X$.

For Statement 2), since $X$ and $Y$ are NSND strategy matrices, Statement 1) implies that every country's safety state is either safe or precarious under both matrices. Hence,
\[
\sum_{l\in\F_i}\mathbb{1}_{\{s_l(X)\geq 0\}}
=
\sum_{l\in\F_i}\mathbb{1}_{\{s_l(Y)\geq 0\}}
=
\operatorname{card}(\F_i),
\]
and $\mathbb{1}_{\{s_j(X)\leq 0\}}=\mathbb{1}_{\{s_j(X)=0\}}$, $\mathbb{1}_{\{s_j(Y)\leq 0\}}=\mathbb{1}_{\{s_j(Y)=0\}}$. It then follows from Definition~\ref{def:preference} that
\begin{equation*}
X\geq_iY
\Longleftrightarrow
\sum_{j\in\A_i}\mathbb{1}_{\{s_j(X)=0\}}
\geq
\sum_{j\in\A_i}\mathbb{1}_{\{s_j(Y)=0\}}.
\end{equation*}
Meanwhile, Statement 1) implies $s_j(X)=x_{jj}$ and $s_j(Y)=y_{jj}$.
Therefore, $X\geq_iY$ if and only if
$\sum_{j\in\A_i}\mathbb{1}_{\{x_{jj}=0\}}\geq\sum_{j\in\A_i}\mathbb{1}_{\{y_{jj}=0\}}$.

\end{proof}

In what follows, we define a matrix transformation that serves as a technical device for constructing a Nash equilibrium. Note that this transformation is not intended as a behavioral rule describing countries' unilateral strategy updates.
\begin{definition}[Preferable adjustment]\label{def:preferable-adjustment}
Given an NSND strategy matrix $X\in \TTheta$ and country $i$, the preferable adjustment of $X$ with respect to $i$, denoted by $\PA^i(X)=\big(\PA^i_{kl}(X)\big)_{n\times n}$, is a strategy matrix constructed as follows:
\begin{enumerate}
\item If $p_i\ge \sum_{j\in \A_i} (x_{jj}+x_{ji})$, let $\PA^i(X)$ be equal to $X$ except for the changes to the following entries:
    \begin{enumerate}
    \item $\PA^i_{ii}(X)=p_i - \sum_{j\in \A_i} (x_{jj}+x_{ji})$;
    \item $\PA^i_{ij}(X)=\PA^i_{ji}(X)=x_{jj}+x_{ji}$ and $\PA^i_{jj}(X)=0$, for any $j\in \A_i$.
    \end{enumerate}
\item If $p_i < \min_{j\in \A_i} (x_{jj}+x_{ji})$, let $\PA^i(X)$ be equal to $X$ except for the changes to the following entries:
   \begin{enumerate}
   \item Randomly pick a country $r$ from  $argmin_{j\in \A_i} (x_{jj}+x_{ji})$. Let $\PA^i_{ii}(X)=0$, $\PA^i_{i r}(X)=\PA^i_{r i}(X)=p_i$, and $\PA^i_{rr}(X)=x_{rr}+x_{ri} - p_i$;
   \item For any $j\in \A_i\setminus \{r\}$, $\PA^i_{ij}(X)=\PA^i_{ji}(X)=0$ and $\PA^i(X)_{jj}=x_{jj}+x_{ji}$.
   \end{enumerate}
\item If $\min_{j\in \A_i} (x_{jj}+x_{ji}) \le p_i < \sum_{j\in \A_i} (x_{jj}+x_{ji})$, let $k=\card(\A_i)$ and index the countries in $\A_i$ as $j_1,\dots,j_k$, with $x_{j_1 j_1}+x_{j_1 i} \le x_{j_2 j_2} + x_{j_2 i}\le \dots \le x_{j_k j_k}+x_{j_k i}$. In this scenario, there exists $m\in \{1,\dots, k-1\}$ such that
\begin{align*}
\sum_{s=1}^m (x_{j_s j_s}+x_{j_s i}) \le p_i < \sum_{s=1}^{m+1} (x_{j_s j_s}+x_{j_s i}).
\end{align*}
Let $\PA^i(X)$ be equal to $X$ except for the changes to the following entries: $\PA^i_{ii}(X)=0$ and,
   \begin{enumerate}
   \item for any $s\in \{1,\dots, m\}$, $\PA^i_{i j_s}(X)=\PA^i_{j_s i}(X)=x_{j_s j_s}+x_{j_s i}$ and $\PA^i_{j_s j_s}(X)=0$;
   \smallskip
   \item $\PA^i_{i j_{m+1}}(X)=\PA^i_{j_{m+1} i}(X)=p_i-\sum_{s=1}^m (x_{j_s j_s} + x_{j_s i})$ and $\PA^i_{j_{m+1} j_{m+1}}(X) = x_{j_{m+1} j_{m+1}} + x_{j_{m+1} i} - \PA^i_{j_{m+1} i}(X)$;
   \smallskip
   \item for any $s\in \{m+2,\dots, k\}$, $\PA^i_{i j_s}(X)=\PA^i_{j_s i}(X)=0$ and $\PA^i_{j_s j_s}(X)=x_{j_s j_s}+x_{j_s i}$.
   \end{enumerate} 
\end{enumerate}
\end{definition}

The intuition underlying the construction of the preferable adjustment can be summarized as follows: the $i$-th row of $\PA^i(X)$ is constructed by reallocating country $i$'s power so as to increase, whenever possible, the number of country $i$'s precarious enemies. The accompanying adjustments to entries outside the $i$-th row preserve the symmetry of $\PA^i(X)$. Consequently, $\PA^i(X)$ remains an NSND strategy matrix, and its $i$-th row represents country $i$’s best response to $X$. The following lemma presents several properties of the preferable adjustment needed for the equilibrium construction. A detailed proof is provided in the earlier conference version of this paper~\cite{zhang2023nash}.

\begin{lemma}[Properties of preferable adjustment]\label{lem: preferable-adjustment-stillNSND-best-response} 
Given any NSND strategy matrix $X\in \TTheta$ and any $i\in \V$, 
\begin{enumerate}
\item $\PA^i(X)$ is also an NSND strategy matrix;
\item for any $Y\in \TTheta$ identical to $X$ except possibly in the $i$-th row, country $i$ prefers $\PA^i(X)$ to $Y$, i.e., $\PA^i(X)\ge_i Y$.
\item $\sum_{j\in \A_i} \mathbb{1}_{\{ \PA^i_{jj}(X)=0\}} \ge \sum_{j\in \A_i} \mathbb{1}_{\{ x_{jj}=0\}}$.
\end{enumerate}
\end{lemma}

With the groundwork laid above, we are ready to prove the existence of a Nash equilibrium as stated in Theorem \ref{thm:existence-NE}.

\textit{Proof of Theorem \ref{thm:existence-NE}}: Given $n$ countries with powers $p_1,\dots,p_n$ and their respective sets of friendly and antagonistic countries $\F_1,\A_1,\dots,\F_n,\A_n$, we construct an iteration process $\{X(t)\}_{t\in \mathbb{N}}$  with $X(0)=\diag(p_1,\dots, p_n)$. By Definition~\ref{def:NSND-strategy-matrix}, $X(0)$ is an NSND strategy matrix. The iteration of $X(t)$ is conducted via the following two procedures.  

\underline{\emph{Procedure 1:}} Given any NSND strategy matrix $X(t)$, randomly pick a pair $(i,j)$ satisfying: 1) $j\in \A_i$; 2) $x_{ii}(t)\neq 0$ and $x_{jj}(t)\neq 0$. Then let $X(t+1)=\PA^i(X(t))$. Repeat the above process until there does not exist such a pair $(i,j)$.

\underline{\emph{Procedure 2:}} Given any NSND strategy matrix $X(t)$, let $\mathcal{B}=\{i\in \V\,|\,x_{ii}(t)=0\}$. Randomly pick an $i\in \mathcal{B}$ such that $\PA^i(X(t))$ satisfies
\begin{align*}
\sum_{j\in \A_i} \mathbb{1}_{\{\PA^i_{jj}(X(t))=0\}}\ge \sum_{j\in \A_i} \mathbb{1}_{\{x_{jj}(t)=0\}} + 1.
\end{align*}
Then let $X(t+1)=\PA^i(X(t))$ and return to Procedure 1. If no such $i\in\mathcal{B}$ exists, then terminate the iteration process.

Since $X(0)$ is an NSND strategy matrix and all the iterations in Procedures~1 and 2 are preferable adjustments, $X(t)$ is NSND for any $t\ge 0$. For any $X\in \TTheta$, define 
\begin{align*}
V(X) = \sum_{i=1}^n \mathbb{1}_{\{x_{ii}=0\}}.
\end{align*}

At any $t$, suppose $X(t+1)$ is obtained from $X(t)$ via a preferable adjustment in Procedure~1 for some $i$, then, according to the conditions for triggering Procedure~1, $x_{ii}(t)\neq 0$ and there exists $j\in \A_i$ such that $x_{jj}(t)\neq 0$. In this case,
\begin{enumerate}
\item if $p_i\ge \sum_{j\in \A_i} (x_{jj}(t)+x_{ji}(t))$, then, by Definition~\ref{def:preferable-adjustment}, we have $x_{ii}(t+1)\ge 0$ and, for any $j\in \A_i$, 
    \begin{align*}
    x_{ij}(t+1) & = x_{ji}(t+1)=x_{jj}(t)+x_{ji}(t),\\
    x_{jj}(t+1) &=0.
    \end{align*}
In addition, for any $k\!\notin\! \A_i\!\cup\! \{i\}$, $x_{kk}(t\!+\!1)\!=\!x_{kk}(t)$. Since $x_{ii}(t)\neq 0$, $x_{ii}(t+1)\ge 0$, $x_{jj}(t+1)=0$ for any $j\in \A_i$, and there exists $j\in \A_i$ such that $x_{jj}(t)\neq 0$, we have
    \begin{align*}
    V&(X(t+1))-V(X(t)) \\
    &=\sum_{j\in \A_i\cup \{i\}}\mathbb{1}_{\{x_{jj}(t+1)=0\}}- \sum_{j\in \A_i\cup \{i\}}\mathbb{1}_{\{x_{jj}(t)=0\}}\ge 1.
    \end{align*}

\item if $p_i < \sum_{j\in \A_i}(x_{jj}(t)+x_{ji}(t))$, then, by Definition~\ref{def:preferable-adjustment},
   \begin{enumerate}
   \item $x_{ii}(t+1)=0$ (while $x_{ii}(t)\neq 0$);
   \item $x_{kk}(t+1)=x_{kk}(t)$ for any $k\notin \A_i \cup \{i\}$.
   \end{enumerate}
In addition, according to Lemma~\ref{lem: preferable-adjustment-stillNSND-best-response}, we have $X(t+1)\ge_i X(t)$, which, according to Statement 2) of Fact~\ref{fact:NSND-properties}, implies that $\sum_{j\in \A_i} \mathbb{1}_{\{x_{jj}(t+1)=0\}}\ge \sum_{j\in \A_i}\mathbb{1}_{\{x_{jj}(t)=0\}}$. Therefore, we have
\begin{align*}
    V&(X(t+1))-V(X(t))\\
    &=\sum_{j\in \A_i} \mathbb{1}_{\{x_{jj}(t+1)=0\}} - \sum_{j\in \A_i}\mathbb{1}_{\{x_{jj}(t)=0\}}\\
    &\quad+\mathbb{1}_{\{x_{ii}(t+1)=0\}}-\mathbb{1}_{\{x_{ii}(t)=0\}} \ge 1.
\end{align*}
\end{enumerate}
Combining 1) and 2), we conclude that, after each iteration via Procedure~1, $V(X(t+1))\ge V(X(t))+1$.

Now we consider the case in which $X(t+1)$ is obtained from $X(t)$ via a preferable adjustment in Procedure~2 for some $i$. Since Procedure~2 requires $x_{ii}(t)=0$ and since $X(t)$ is an NSND strategy matrix, we have $p_i=\sum_{j\in \A_i} x_{ij}(t)=\sum_{j\in \A_i} x_{ji}(t)$. Suppose $x_{jj}(t)=0$ for any $j\in \A_i$, by Definition~\ref{def:preferable-adjustment}, 
\begin{align*}
x_{ij}(t+1) = x_{ij}(t) = x_{ji}(t) = x_{ji}(t+1)
\end{align*}
and $x_{ii}(t+1)=0$. In this case, 
\begin{align*}
\sum_{j\in \A_i}\mathbb{1}_{\{x_{jj}(t+1)=0\}} = \sum_{j\in \A_i}\mathbb{1}_{\{x_{jj}(t)=0\}},
\end{align*}
which contradicts the rule of Procedure~2. Therefore, there must exist $j\in \A_i$ such that $x_{jj}(t)>0$, which leads to 
\begin{align*}
p_i<\sum_{j\in \A_i} (x_{jj}(t)+x_{ji}(t)).
\end{align*}
By Definition~\ref{def:preferable-adjustment}, $x_{ii}(t+1)=0$ and $x_{kk}(t+1)=x_{kk}(t)$ for any $k\notin \A_i\cup \{i\}$. Therefore, 
\begin{align*}
V( & X(t+1))-V(X(t))\\
&=\sum_{j\in \A_i} \mathbb{1}_{\{x_{jj}(t+1)=0\}}-\sum_{j\in \A_i} \mathbb{1}_{\{x_{jj}(t)=0\}}\ge 1.
\end{align*}

Based on the above discussions for both Procedure~1 and Procedure~2, we conclude that after each iteration of $X(t)$, $V(X(t))$ increases by at least 1. Moreover, since $V(X(t))$ is upper bounded by $n$, the iteration $X(t)$ must terminate at some finite time step $T$.

Now we proceed to prove that $X(T)$ must be a Nash equilibrium of the static power-allocation game. For any $i$, suppose $x_{ii}(T)\neq 0$. Since Procedure~1 can no longer be activated, we have $x_{jj}(T)=0$ for any $j\in \A_i$. According to Statement 1) of Fact~\ref{fact:NSND-properties}, all of $i$'s antagonistic countries are already precarious. That is, $i$ cannot further increase the number of its precarious antagonistic countries. Moreover, since $X(T)$ is NSND, $i$ and all of $i$'s friendly countries are either safe or precarious under $X(T)$. Therefore, the $i$-th row of $X(T)$ is already country $i$'s best response to $X(T)$. Suppose $x_{ii}(T)=0$. Since Procedure~2 cannot be activated, the preferable adjustment of $X(T)$ for $i$ does not increase the number of $i$'s precarious antagonistic countries. Moreover, since $X(T)$ is NSND, $i$ and all of $i$'s friendly countries are either safe or precarious under $X(T)$. Therefore, the $i$-th row of $X(T)$ is already country $i$'s best response to $X(T)$. 

Now we have proved that each row of $X(T)$ is the corresponding country's best response to $X(T)$. That is, $X(T)$ is a Nash equilibrium. Finally, since $X(T)$ is an NSND strategy matrix, no country is in danger under $X(T)$ and no country supports any of its friendly countries, i.e., $x_{ij}(T)=0$ for any $i,\,j$ such that $j\in \F_i\setminus\{i\}$. This concludes the proof.\hfill$\blacksquare$

Theorem~\ref{thm:existence-NE} shows that, for any static power-allocation game defined in Section~II-B, there exists a special Nash equilibrium in which no country is in danger without providing support to others. It is worth mentioning that Theorem~\ref{thm:existence-NE} guarantees the existence of an NSND Nash equilibrium, but does not imply that all Nash equilibria are NSND. This NSND equilibrium also admits a meaningful real-world interpretation. With positive support allocations between distinct friendly countries interpreted as active alliance commitments, this equilibrium shares structural similarities with the real-world political agenda known as the ``Non-Aligned Movement''\footnote{We do not claim historical equivalence, but a structural similarity in incentive compatibility.}~\cite{mital2016non}.

\begin{remark}\label{remark:Example}
    To illustrate the existence of multiple Nash equilibria and the distinction between NSND and non-NSND equilibria, consider the example in Fig.~\ref{fig:NE_example}. The network consists of six countries with powers $p_1=10$, $p_2=20$, $p_3=27$, $p_4=15$, $p_5=17$, and $p_6=20$. The figure presents two power-allocation profiles, both of which constitute Nash equilibria. The equilibrium in Fig.~\ref{fig:NE_example}A belongs to the NSND class identified in Theorem~\ref{thm:existence-NE}, whereas the one in Fig.~\ref{fig:NE_example}B is a non-NSND equilibrium. In the latter equilibrium, countries $5$ and $6$ are in danger.
    \begin{figure}[t]
        \centering
        \includegraphics[width=\columnwidth]{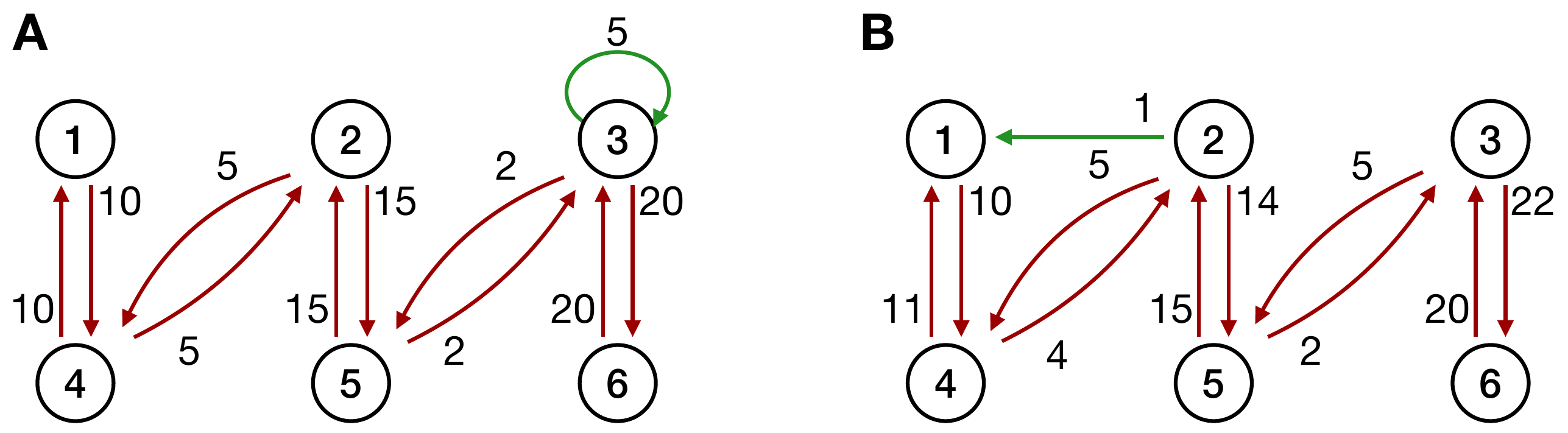}
        \caption{Two Nash equilibria for the six-country example discussed in Remark~\ref{remark:Example}. The nodes represent countries, with red and green edges indicating antagonistic and friendly relations, respectively. The number on each directed arrow indicates the power allocated to attack or support the target country, depending on whether the relation is antagonistic or friendly.}
        \label{fig:NE_example}
    \end{figure}
\end{remark}

\subsection{Extreme-case analysis: Nash equilibria in a fully antagonistic world
}
In this subsection, we consider an extreme case where all countries are mutually antagonistic. The following theorem determines the maximum number of countries that can remain safe in such a world.

\begin{theorem}[Number of safe countries in a fully antagonistic world]\label{thm:number-NE}
Consider the static power-allocation game defined in Section~\ref{subsection:static-model-setup} with $n\ge 3$ countries. Assume that $p_1\leq p_2\leq \cdots\leq p_n$. Suppose that each country has a positive power and all the countries are hostile to each other, i.e., $\A_i=\V\setminus\{i\}$ for any $i\in \V$.

\begin{enumerate}
    \item Given any power vector $p\in \mathbb{N}^n$ and at any Nash equilibrium $X$, there are at most $n-1$ safe countries. Namely, it is impossible for every country to be safe.
    \item If $p_{n}>\sum_{i=1}^{n-1} p_i$, then there is only one safe country at any Nash equilibrium, which is country $n$. 
    \item If $p_{n}\leq\sum_{i=1}^{n-1} p_i$, then there always exists a Nash equilibrium with no safe country.
\end{enumerate}
\end{theorem}

\begin{proof}
We first prove Statement 1) by contradiction. Suppose that all countries are safe at some Nash equilibrium $X\in \TTheta$, which implies that no country could be in a more preferred situation by reallocating its power. Then, for any countries $i,j\in\V$ with $i\neq j$, country $j$ cannot make country $i$ not safe by allocating all of its power to attacking country $i$. That is,
    \begin{align*}
       p_{j}\!-\!x_{ji} <\!\sum_{s\in\F_i} x_{s i}+\!\!\sum_{s\in\A_i} x_{i s}\!-\!\!\sum_{s\in\A_i} x_{s i}
        =x_{ii}\!+\!\sum_{s\neq i}(x_{i s}\!-\!x_{s i}). 
    \end{align*}
Let $j=i+1$. Here we treat country $n+1$ as country $1$ and treat country $0$ as country $n$. By summing the left-hand side of the above inequality over all $i\in \V$, we obtain
    \begin{align*}
        \sum_{i\in\V}p_i-\!\sum_{i\in\V}x_{ii-1}\!<\!\sum_{i\in\V}x_{ii}\!+\!\sum_{i\in\V}\sum_{s\neq i}(x_{i s}\!-\!x_{s i})=\!\sum_{i\in\V}x_{ii}.
    \end{align*}
    Hence,
    \begin{align*}
        \sum_{i\in\V}\sum_{j\in\V}x_{ij}=\sum_{i\in\V}p_i<\sum_{i\in\V}(x_{ii-1}+x_{ii}),
    \end{align*}
    which leads to a contradiction and thus proves Statement~1).

The proof of Statement 2) is straightforward. If $p_{n}>\sum_{i=1}^{n-1} p_i$, no matter how other countries allocate their powers, it is impossible to make country $n$ unsafe. Meanwhile, given the power-allocation strategies of other countries, country $n$ can make all countries except itself unsafe by assigning power $p_i$ on attacking country $i$ for any $i\in \V\setminus\{n\}$. Therefore, at any Nash equilibrium, there is only one safe country, which is country $n$.
    
Now we prove Statement 3). We first show by induction that, when $n\ge 3$ and $\sum_{i\in \V}p_i$ is an even integer, there exists a Nash equilibrium such that every country is precarious and thus no one is safe. When $n=3$, we construct a strategy matrix $X^*=(x^*_{ij})_{3\times 3}$ as follows:
    \begin{equation}\label{eq:spec-net-induction}
        \begin{aligned}
        x^*_{11}&\!=\!x^*_{22}\!=\!x^*_{33}\!=\!0;\,\, &&x^*_{12}\!=\!x^*_{21}\!=\!\frac{p_1+p_2-p_3}{2};\\
        x^*_{13}&\!=\!x^*_{31}\!=\!\frac{p_1+p_3-p_2}{2}; &&x^*_{23}\!=\!x^*_{32}\!=\!\frac{p_2+p_3-p_1}{2}.
        \end{aligned}
    \end{equation}
Since $p_i\le \sum_{j\neq i}p_j$ for any $i\in \V$, $X^*$ is entry-wise nonnegative. As $p_1+p_2+p_3$ is even, any off-diagonal entry of $X^*$ is an integer. Moreover, $\sum_{j\in \V}x^*_{ij}=p_i$ for any $i\in \V$, so $X^*$ is a feasible strategy matrix. Moreover, one can check that every country is precarious in this case, which is already the most preferred situation for every country. Therefore, such a strategy matrix $X^*$ is a Nash equilibrium. 
   
Suppose that the induction hypothesis holds for some $n\ge 3$. In the case with $n+1$ countries, our discussion is split into two cases:
    
\emph{Case 1:} $p_{n+1}+2p_{n-1}>\sum_{i=1}^n p_i$. We can explicitly construct a Nash-equilibrium strategy matrix $X^*=(x^*_{ij})_{(n+1)\times (n+1)}$ such that no country is safe. This matrix $X^*$ is constructed as follows:
    \begin{align*}
        x^*_{i,n+1}&=x^*_{n+1,i}=p_i,\quad \text{for any }i<n-1;\\
        x^*_{n-1,n}&=x^*_{n,n-1}=\frac{\sum_{i=1}^n p_i - p_{n+1}}{2};\\
        x^*_{n,n+1}&=x^*_{n+1,n}\!=\!\frac{p_{n+1}\!+\!2p_{n}\!-\!\!\sum_{i=1}^{n} p_i}{2};\\
        x^*_{n-1,n+1}&=x^*_{n+1,n-1}\!=\!\frac{p_{n+1}\!+\!2p_{n-1}\!-\!\!\sum_{i=1}^{n} p_i}{2};\\
        x^*_{ij}&=x^*_{ji}=0, \quad\text{otherwise}.
    \end{align*}
Since $p_{n+1}\le \sum_{i=1}^n p_i$, each entry of $X^*$ is nonnegative. Since $2x^*_{n,n-1}+2p_{n+1}=\sum_{i=1}^{n+1}p_i$ is even, $x^*_{n,n-1}=x^*_{n-1,n}$ is an integer. Since $2x^*_{n,n+1}+\sum_{i=1}^{n+1}p_i=2(p_n+p_{n+1})$ is even, $x^*_{n,n+1}=x^*_{n+1,n}$ is an integer. Similarly, $x^*_{n-1,n+1}=x^*_{n+1,n-1}$ is an integer. Hence, each entry of $X^*$ is an integer. In addition, one can check that $\sum_{j=1}^{n+1}x^*_{ij}=p_i$ for any $i\in \{1,\dots, n,n+1\}$. That is, $X^*$ is a feasible strategy matrix. Moreover, since $X^*$ is symmetric with zero diagonal and all relations are antagonistic, every country is precarious under $X^*$, which in turn implies that $X^*$ is a Nash equilibrium, where no country is safe.

\emph{Case 2:} $p_{n+1}-(p_n-p_{n-1})\leq \sum_{i=1}^{n-2}p_i$. Then there exists $1\leq k\leq n-2$ such that 
    \begin{align*}
        \sum_{i=1}^{k-1}p_i< p_{n+1}-(p_n-p_{n-1})\leq \sum_{i=1}^{k}p_i.
    \end{align*}
Here we adopt the shorthand $\sum_{i=1}^0 p_i=0$ if needed. We construct an auxiliary system with $n$ mutually antagonistic countries whose powers $p_1^*,\dots, p_n^*$ are given as follows:
    \begin{align*}
        p_i^*&=0,\quad \text{for any }i<k;\\
        p_k^*&=\sum_{j=1}^k p_j-p_{n-1}+p_n-p_{n+1};\\
        p_i^*&=p_i,\quad \text{for any }k<i\le n-1;\\
        p_n^*&=p_{n-1}.
    \end{align*}
Since $p_{n+1}-(p_n-p_{n-1})\leq \sum_{i=1}^{n-2}p_i$, for any $i\in\{1,\dots, n\}$, $p_i^*$ is a nonnegative integer. In addition, 
    \begin{align*}
        \sum_{i=1}^np^*_i=\sum_{j=1}^{n-1} p_j + p_n -p_{n+1} = \sum_{i=1}^{n+1}p_i -2p_{n+1}.
    \end{align*}
Therefore, $\sum_{i=1}^n p_i^*$ is even. The order $p_1\le p_2\le\dots\le p_n$ implies $p^*_1\le p^*_2\le\dots\le p^*_n$. Furthermore,
\begin{align*}
2p^*_n-\sum_{j=1}^n p_j^* = p_{n+1}-(p_n-p_{n-1})-\sum_{j=1}^{n-2}p_j\le 0,
\end{align*}
and hence $2p^*_i\le \sum_{j=1}^n p_j^*$ for all $i\in\{1,\dots,n\}$. Thus, all conditions for the induction hypothesis are satisfied. Consequently, for this auxiliary system, there exists a Nash equilibrium $X^*$ under which every country is precarious. 

For the original $(n+1)$-country system with powers $p_1,\dots, p_{n+1}$, construct a strategy matrix $X=(x_{ij})_{(n+1)\times (n+1)}$ whose first $n$ rows and columns coincide with $X^*$. Let
\begin{align*}
    x_{n+1,i}=x_{i,n+1}=p_i-p_i^*,\quad\text{for any }i\in \{1,\dots,n\},
\end{align*}
and $x_{n+1,n+1}=0$. One can verify that $X$ is feasible, and that every country is precarious under $X$. Hence, $X$ is a Nash equilibrium. This completes the induction proof that, when the sum of all countries’ powers is even, there exists a Nash equilibrium such that all countries are precarious.

If the sum of all countries’ power is odd, we have $p_n\leq \sum_{i=1}^{n}p_i-1$. Take countries $n-2,n-1,n$, and temporarily reduce their power by 1. Now the countries' total power is even. We have proved that, in this case, there exists a Nash equilibrium $X$ such that all countries are precarious. Then we restore the powers previously removed to the three countries and add $1$ to $x_{n-2,n-1}, x_{n-1,n}, and\ x_{n,n-2}$ respectively. The new strategy matrix remains feasible. Moreover, it is still a Nash equilibrium since all countries are precarious. This completes the proof of Statement 3).
\end{proof}
Taken together, Theorem~\ref{thm:number-NE} reveals how adversarial network structures and power distributions jointly shape equilibrium security patterns. Mutual hostility imposes a fundamental limitation: all countries cannot be safe simultaneously at equilibrium, regardless of how power is distributed. However, the distribution of power determines the form of equilibrium security. When one country possesses more power than all the others combined, security is concentrated exclusively in that country; when no dominant country exists, an equilibrium may instead arise in which all countries are precarious. Thus, power concentration enables individual security at the expense of others, whereas a more balanced distribution prevents domination but leads to widespread vulnerability. These results provide a network-game-theoretic perspective on the tension between power concentration and collective security in adversarial international environments.

\section{Dynamic model: coevolution of countries' powers and strategies}
Intuitively, a country in a safe environment is more likely to prosper, whereas a country in danger tends to decline. This pattern is supported by empirical evidence provided in Section~IV-A (see the paragraph ``Post-war power dynamics''), which relates countries’ Nash-equilibrium states to their subsequent GDP growth. These observations motivate us to study how signed network structures shape the evolution of nations' powers and, when conflicts are inevitable, which configurations of signed networks better foster global prosperity. To address these questions, we extend the static power-allocation game to a dynamic model of the coevolution between countries’ powers and power-allocation strategies, referred to as the \emph{\textbf{co}evolutionary \textbf{p}ower-\textbf{s}trategy dynamics} (COPS dynamics).

\subsection{Model setup and notation}\label{subsection:COPS}

\emph{Network, power allocation, and countries' state:} The COPS dynamics adopt a similar network setup to the static power-allocation game. Consider $n$ countries embedded in a signed network, where $\F_i$ and $\A_i$ denote the sets of friendly and antagonistic neighbors for any country $i\in \V$, respectively. Similar to the notation used in Section~\ref{subsection:static-model-setup}, denote by $X=(x_{ij})_{n\times n}$ the matrix of all the countries' power-allocation strategies, which evolves with time.  Let $K\in \mathbb{N}_+$ be a prescribed upper limit of any country's power, and let
\begin{align*}
    \OOmega=\Big{\{} X\in \mathbb{N}^{n\times n}\,\Big|\,&\sum_{j=1}^n x_{ij}\le K\text{ for any }i\in \V,  \\ 
    &\text{and }x_{ij}=0 \text{ for any } j\notin \A_i\cup\F_i \Big{\}}
\end{align*}
be the set of all possible strategy matrices. Given a strategy matrix $X\in \OOmega$, the state of any country $i\in \V$, denoted by $s_i(X)$, is defined by Definition~\ref{def:state} and is classified into three types: safe, precarious, or in danger, depending on whether $s_i(X)$ is positive, zero, or negative, respectively. For any subset of countries $\tilde{\V}\subseteq \V$, let $N_p(\tilde{\V},X)$ be the number of precarious countries in the set $\tilde{\V}$ under the strategy matrix $X$. The following lemma will be frequently used in this section. Its proof is provided in Appendix~\ref{proof:lem:impossible-all-danger}.

\begin{lemma}\label{lem:impossible-all-danger}
    Every strategy matrix $X\in \OOmega$ satisfies $\sum_{i\in \V} s_i(X)\ge 0$. As a result, there always exists a safe or precarious country. Moreover, if at least one country is in danger under $X$, then there exist $i\in \V$ and $j\in \A_i$ such that $s_i(X)\ge 0$, $s_j(X)<0$, and $x_{ij}>0$. That is, whenever the set of countries in danger is nonempty, at least one country in danger is attacked by a country not in danger.
\end{lemma}

\emph{Specified utility function:} For the dynamic model, we define a specific utility function. For each country $i\in\V$, its utility under a strategy matrix $X$ is given by
\begin{align*}
    u_i(X)&=\sum_{j=1}^n u_{ji}(X)+n\mathbb{1}_{\{s_i(X)\ge0\}},\quad \text{where}\\
    u_{ji}(X) & = 
       \begin{cases}
       1, & \text{if } j \in \F_i\text{ and }s_j(X) \ge 0, \\
       & \text{or if } j \in \A_i\text{ and }s_j(X) \leq 0, \\
       0, & \text{otherwise},
\end{cases}
\end{align*}
which represents the utility that country $i$ obtains from the state of country $j$. Consistent with the preference axioms in Section~\ref{subsection:static-model-setup}, the structure of $u_i(X)$ indicates that ``being not in danger'' is the first priority of each country; beyond self-preservation, a country aims to maximize the number of friendly countries that are not in danger and antagonistic countries that are not safe.

\emph{Coevolution of powers and strategies:} We model the COPS dynamics as an asynchronous discrete-time stochastic process. The system starts with an initial power vector $p(0)=\big(p_1(0),\dots,p_n(0)\big)^{\top}\in \{0,1,\dots,K\}^n$ and an initial strategy matrix $X(0)\in \OOmega$ satisfying $X(0)\mathbf{1}_n=p(0)$. At each time $t\in \mathbb{N}_+$, a country $i_t\in \V$ is randomly activated and updates its power and power-allocation strategy, while any other country's power and strategy remain unchanged, i.e., $p_j(t)=p_j(t-1)$ and $x_{jk}(t)=x_{jk}(t-1)$ for any $j\neq i_t$ and any $k\in \V$. The power of the activated country $i_t$ is updated based on its state at time $t-1$ as follows:
\begin{align*}
    p_{i_t}(t)=
    \begin{cases}
    \displaystyle \min\{p_{i_t}(t-1)+1,K\},&\,\, \text{if }s_{i_t}(X(t-1))>0,\\
    \displaystyle p_{i_t}(t-1),&\,\, \text{if }s_{i_t}(X(t-1))=0,\\
    \displaystyle \max\{p_{i_t}(t-1)-1,0\},&\,\, \text{if } s_{i_t}(X(t-1))<0.
    \end{cases}
\end{align*}
That is, $p(t)$ is determined by $X(t-1)$ and $i_t$. 

Given the updated power $p_{i_t}(t)$, country $i_t$ uniformly randomly selects a power-allocation vector from a candidate set obeying the total-power constraint. As a result, the updated strategy matrix $X(t)$ is uniformly randomly chosen from a candidate set $\bfU_{i_t}\big( X(t-1) \big)$, which is a subset of the feasible set
\begin{align*}
    \OOmega_0 = \Big{\{} X\in \OOmega\,\Big|\, & x_{jk}=x_{jk}(t-1), \forall\,j\neq i_t,\,\forall\, k\in\V;\\
    & X\mathbf{1}_n=p(t)\Big{\}}.
\end{align*}
In principle, country $i_t$ seeks to increase or at least preserve its utility, if possible. Accordingly, the candidate set $\bfU_{i_t}\big(X(t-1)\big)$ is defined by the following four rules, which together specify a conflict-aware and boundedly rational update process.

\textbf{Rule 1:} $\bfU_{i_t}\big(X(t-1)\big) \subseteq \OOmega_1$, where $\OOmega_1$ is a subset of $\OOmega_0$ such that, for any $j\neq i_t$,
\begin{enumerate}
    \item $x_{i_tj}\le x_{i_t j}(t-1)$, if $u_{ji_t}(X(t-1))=1$;
    \item $x_{i_t j}\le x_{i_t j}(t-1)+|s_j(X(t-1))|$, if $u_{j i_t}(X(t-1))=0$.
\end{enumerate}

\textbf{Rule 2:} If $\OOmega_2=\{X\in \OOmega_1\,|\, u_{i_t}(X)>u_{i_t}(X(t-1))\}$ is non-empty, then $\bfU_{i_t}\big(X(t-1)\big)=\OOmega_2$;

\textbf{Rule 3:} If $\OOmega_2$ is empty and
\begin{align*}
    \OOmega_3 = \Big{\{} X\in \OOmega_1\,\Big|\, & u_{i_t}(X)=u_{i_t}(X(t-1));\\
    & x_{i_t j}\le x_{i_t j}(t-1),\,\,\forall\, j\in \A_{i_t};\\
    & x_{i_t j}=x_{i_t j}(t-1),\,\, \forall\, j\in \F_{i_t}\setminus \{i_t\}\Big{\}}
\end{align*}
is non-empty, then $\bfU_{i_t}\big(X(t-1)\big)=\OOmega_3$. If $\OOmega_3$ is empty but 
\begin{align*}
    \OOmega_4=\{X\in \OOmega_1\,|\, u_{i_t}(X)=u_{i_t}(X(t-1))\}
\end{align*}
is non-empty, then $\bfU_{i_t}\big(X(t-1)\big)=\OOmega_4$;

\textbf{Rule 4:} If the sets $\OOmega_2$, $\OOmega_3$, $\OOmega_4$ are all empty, country $i_t$ randomly chooses a feasible strategy obeying Rule 1, i.e., $\bfU_{i_t}\big(X(t-1)\big)=\OOmega_1$.

These four rules describe an asynchronous, boundedly rational revision process. The intuition underlying them is as follows. Rule~1 prevents country $i_t$ from unnecessarily increasing its power allocation to other countries, thereby avoiding wasteful resource expenditures. Within $\OOmega_1$, Rule~2 captures a natural better-response behavior: country $i_t$ adopts a new strategy whenever it can strictly improve its utility. If no utility-improving adjustment is available and a strategy change can only preserve the current utility, Rule~3 introduces an additional preference for de-escalation by selecting strategies that avoid unnecessary increases in conflict intensity. Finally, when no utility-improving or utility-preserving strategy is available, Rule~4 provides a feasible fallback by allowing country $i_t$ to choose randomly from $\OOmega_1$.

\begin{figure}
    \centering
    \includegraphics[width=1\linewidth]{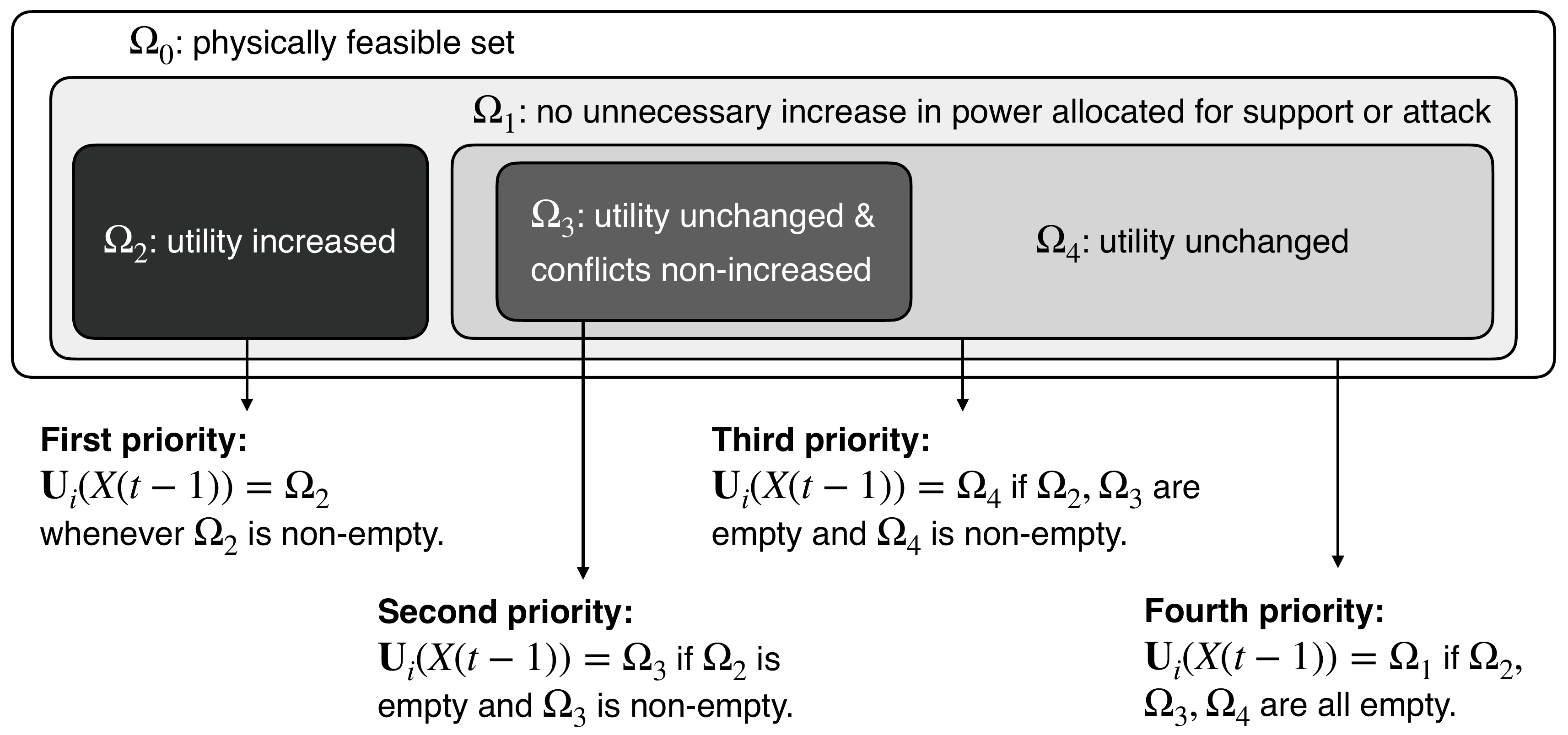}
    \caption{A Venn diagram illustrating Rules 1--4 for the updates of countries' power-allocation strategies.}
    \label{fig:dynamic-rule}
\end{figure}

The way the candidate set $\bfU_{i_t}\big(X(t-1)\big)$ is constructed defines the stochastic update rule of the strategy matrix $X(t)$. The relations between Rules 1--4 and the sets $\OOmega_0$, $\OOmega_1$, $\OOmega_2$, $\OOmega_3$, $\OOmega_4$ are illustrated by Fig.~\ref{fig:dynamic-rule}. Rules 1--4 immediately lead to the following lemma that will be repeatedly applied.
\begin{lemma}\label{lem:safe-prec-utility-nondecrease}
    For any $X\in \OOmega$ and any $i\in \V$ such that $s_i(X)\ge 0$, the candidate set $\bfU_i(X)$ defined by Rules~1--4 satisfies
    \begin{enumerate}
        \item either $\bfU_i(X)=\OOmega_2$ or $\bfU_i(X)=\OOmega_3$;
        \item $u_i(X^+)\ge u_i(X)$ and $s_i(X^+)\ge  0$ for any $X^+\in \bfU_i(X)$.
    \end{enumerate}
\end{lemma}
\begin{proof}
Given the strategy matrix $X$, suppose country $i$, with $s_i(X)\ge 0$, is activated to update its power and its power-allocation strategy. Let $p_i$ be its current power under $X$, i.e., $p_i=(X\mathbf{1}_n)_i$, and let $p_i^+$ be its updated power. Since $s_i(X)\ge 0$, we have $p_i^+\ge p_i$. In this case, country $i$ can ensure that its utility does not decrease by keeping all allocations to other countries unchanged and assigning $x_{ii}+p_i^+-p_i$ to itself. Hence $\OOmega_3$ in Rule~3 is non-empty. According to Rules~1--4, 
\begin{align*}
    \bfU_i(X)=
    \begin{cases}
        \OOmega_2\quad &\text{if }\OOmega_2\text{ is non-empty};\\
        \OOmega_3\quad &\text{otherwise}.
    \end{cases}
\end{align*}
By the definition of $\OOmega_2$ and $\OOmega_3$, $u_i(X^+)\ge u_i(X)$ for any $X^+\in \bfU_i(X)$. This in turn implies $s_i(X^+)\ge 0$, otherwise $u_i(X^+)< n < u_i(X)$. This concludes the proof.
\end{proof}

\emph{System state, trajectory, and equilibrium:} As implied by Rules 1--4, $X(t)$ alone characterizes the state of the COPS dynamics at time $t$, and $p(t)$ can be inferred via $p(t)=X(t)\mathbf{1}_n$. Let $\{i_t\}_{t\in\mathbb{N}_+}$ be the sequence of activated countries. Given $X(0)$ and $\{i_t\}_{t\in\mathbb{N}_+}$, the trajectory $\{X(t)\}_{t\in\mathbb{N}}$ forms a stochastic process, where $X(t)$ depends on $X(t-1)$, $i_t$, and also the random choice of a strategy matrix from the candidate set $\bfU_{i_t}\big(X(t-1)\big)$.

A strategy matrix $X^*\in \OOmega$ is an equilibrium of the COPS dynamics if $\bfU_i(X^*)=\{X^*\}$ for any $i\in \V$.  

\begin{remark}\label{remark:dyn-eq-is-static-NE}
Suppose $X^*$ is an equilibrium of the COPS dynamics. According to the update rule for countries' powers, if $s_i(X^*)<0$ for some $i$, then the power of country $i$, i.e., $\sum_{k\in \V}x_{ik}^*$, must be zero. In addition, for any $i\in \V$ with $s_i(X^*)\ge 0$, according to Rules 1--3, $\bfU_i(X^*)=\{X^*\}$ implies that country $i$ cannot increase its utility by changing its power-allocation strategy. Note that the utility function $u_i(\cdot)$ defined in Section III-A is compatible with the preference axiom given by Definition~\ref{def:preference} in Section~II. Therefore, $X^*$ is also a Nash equilibrium of the static power-allocation game.    
\end{remark}

\subsection{Convergence analysis}

The core objective of this subsection is to establish the following theorem on the almost-sure convergence of the COPS dynamics to an equilibrium in finite time, which is also a Nash equilibrium of the corresponding SPA game.

\begin{theorem}[Convergence of the dynamic process]\label{thm:convergence} Consider the COPS dynamics with any initial strategy matrix $X(0)\in \OOmega$ and the initial power $p(0)=X(0)\mathbf{1}_n$. The trajectory $\{X(t)\}$ almost surely reaches an equilibrium $X^*$ in finite time. 
\end{theorem}

This theorem justifies the COPS dynamics as a well-posed model and enables us to further study how its equilibria change with model parameters such as the signed network structures, which will be investigated numerically in the next section.

We prove Theorem~\ref{thm:convergence} by reducing it to a finite-time reachability problem using the following lemma:
\begin{lemma}\label{lem:Markev}
Consider the COPS dynamics defined in Section~\ref{subsection:COPS}. Suppose that, starting from any $X(0)\in\OOmega$, there exists a feasible trajectory $\{X(t)\}$ that reaches an equilibrium in finite time. Then, starting from any $X(0)\in\OOmega$, $X(t)$ almost surely reaches an equilibrium in finite time.
\end{lemma}
Since $\OOmega$ is finite and the state-transition probabilities depend only on $X(t)$, the process $\{X(t)\}$ is a finite-state Markov chain. The proof of Lemma~\ref{lem:Markev} then follows a standard argument that every state of the Markov chain admits a directed path to an absorbing state, see Lemma~7 in \cite{WM-JMH-GC-FB-FD:24} for a similar result. 

According to Lemma~\ref{lem:Markev}, in order to prove the almost-sure convergence of the COPS dynamics, it suffices to construct, from every initial matrix, a feasible finite-length trajectory that reaches an equilibrium. The trajectory constructed in this subsection involves repeated activations of safe or precarious countries. These updates satisfy the following monotonicity property. The proof is provided in Appendix~\ref{proof:lem:precarious number}.
\begin{lemma}\label{lem:precarious number}
Consider the COPS dynamics defined in Section~\ref{subsection:COPS}. For any $X\in \OOmega$ and for any $i\in \V$ such that $s_i(X)\ge 0$, the inequalities
\begin{align*}
    N_p(\V\setminus\! \{i\},X^+)-N_p(\V\setminus\!\{i\},X)&\ge  u_i(X^+)-u_i(X)\\
    N_p(\V,X^+)-N_p(\V,X)&\ge 0
\end{align*}
hold for any $X^+\in \bfU_i(X)$, where $\bfU_i(X)$ is defined by Rules~1--4 in Section~\ref{subsection:COPS}. 
\end{lemma}

Now we define two sets that serve as intermediate targets along the trajectory we construct:
\begin{align}
   \bfS_1=\Big{\{} & X\in \OOmega\,\Big|\, N_p(\V,\tilde{X}(t))\equiv N_p(\V,\tilde{X}(0)),\, \forall\, t\in \mathbb{N}, \notag \\
    & \text{for any }\{i_t\} \text{ and }\{\tilde{X}(t)\} \text{ such that }\tilde{X}(0)=X \label{eq:S1} \\
    &\text{and }s_{i_t}(\tilde{X}(t-1))\ge 0,\, \forall\, t\in \mathbb{N}_+\Big{\}}.\notag \\
     \bfS_2 =\Big{\{} & X\in \bfS_1 \,\Big|\, \sgn(s_i(\tilde{X}(t)))\equiv \sgn(s_i(\tilde{X}(0))), \notag \\
    &\forall\, i\in\V, t\in\mathbb{N}_+, \text{ for any }\{i_t\}\text{ and }\{\tilde{X}(t)\}\text{ such } \label{eq:S2}\\
    &\text{that }\tilde{X}(0)\!=\! X \text{ and }s_{i_t}\big( \tilde{X}(t-1) \big)\!=\!0,\,\, \forall\, t\!\in\! \mathbb{N}_+\Big{\}}.\notag
\end{align}
Here $\bfS_1$ is the set of strategy matrices from which the number
of precarious countries remains unchanged under any sequence of
updates involving only safe or precarious countries, while $\bfS_2$ is the subset of $\bfS_1$ from which every update sequence involving only precarious
countries preserves each country's safety state. Their reachability and invariance properties are given by the following two lemmas, with their proofs provided in Appendices~\ref{proof:lem:S1} and~\ref{proof:lem:S2}, respectively.

\begin{lemma}\label{lem:S1}
    For any initial state $X(0)\in\OOmega$ satisfying
    $N_p(\V,X(0))>0$, there exists a feasible finite trajectory,
    along which only safe or precarious countries are activated,
    that reaches $\bfS_1$. For any $X\in\bfS_1$, any safe country can update its strategy only according to Rule~3, and such an update preserves every country’s safety state.
\end{lemma}

\begin{lemma}\label{lem:S2}
    For any initial state $X(0)\in\bfS_1$ satisfying
    $N_p(\V,X(0))>0$, there exists
    a feasible finite trajectory, along which only precarious
    countries are activated, that reaches $\bfS_2$. For any $X\in\bfS_2$, every update by a precarious country preserves every country’s safety state.
\end{lemma}

Lemmas~\ref{lem:S1} and~\ref{lem:S2} concern initial states with $N_p(\V,X(0))>0$, while the case when $N_p(\V,X(0))=0$ is addressed by the lemma below. Its proof is provided in Appendix~\ref{proof:lem:no-precarious-initially}.
\begin{lemma}\label{lem:no-precarious-initially}
    Consider the COPS dynamics with any initial state $X(0)$ such that $N_p(\V,X(0))=0$. There exists an activation sequence $\{i_t\}$ and a compatible trajectory $\{X(t)\}$ that reaches, at some finite time $T\in\mathbb{N}$, either an equilibrium or a state satisfying $N_p(\V,X(T))>0$.  
\end{lemma}

Lemma~\ref{lem:no-precarious-initially} reduces the remaining analysis to initial states with $N_p(\V,X(0))>0$. With the groundwork laid above, we now construct a finite trajectory consisting of four phases, as specified in Algorithm~\ref{alg:trajectory}.

\begin{algorithm}
\caption{Construction of a feasible trajectory}
\label{alg:trajectory}
\begin{algorithmic}[1]
\Require An initial strategy matrix $X(0)\in\OOmega$ satisfying
$N_p(\V,X(0))>0$

\Statex
\State \textbf{Phase 1: Reach $\bfS_1$}
\State Follow a feasible finite trajectory guaranteed by
Lemma~\ref{lem:S1}, activating only safe or precarious countries,
until it first reaches $\bfS_1$; denote this time by $T_1$.

\Statex
\State \textbf{Phase 2: Reach $\bfS_2$}
\State Starting from $X(T_1)$, follow a feasible finite trajectory guaranteed by Lemma~\ref{lem:S2}, activating only precarious countries, until it first reaches $\bfS_2$; denote this time by $T_2$.

\Statex
\State \textbf{Phase 3: Eliminate attacks by precarious countries
on non-precarious enemies}
\While{$\exists\, i\in\V, j\in\A_i$ such that $s_i(X)=0$, $s_j(X)\neq 0$, and $x_{ij}>0$}
    \State Randomly pick such a pair $(i,j)$ and let $x_{ij}\gets x_{ij}-1,\quad x_{ii}\gets x_{ii}+1$
\EndWhile
\State Let $T_3\geq T_2$ denote the time at which Phase~3 terminates

\Statex
\State \textbf{Phase 4: Increase the powers of safe countries and
eliminate their attacks on non-precarious enemies}
\While{$\exists\, i\in\V$ such that $s_i(X)>0$ and $p_i<K$}
    \State Randomly pick such an $i$ and let $x_{ii}\gets x_{ii}+1$
\EndWhile
\While{$\exists\, i\in\V, j\in\A_i$ such that $s_i(X)>0$, $s_j(X)\neq 0$, and $x_{ij}>0$}
    \State Randomly pick such a pair $(i,j)$ and let $x_{ij}\gets x_{ij}-1,\quad x_{ii}\gets x_{ii}+1$
\EndWhile
\State Let $T_4\geq T_3$ denote the time at which Phase~4 terminates

\Statex
\State \Return the resulting strategy matrix $X(T_4)$
\end{algorithmic}
\end{algorithm}

To finalize the proof of Theorem~\ref{thm:convergence}, it now suffices to show that Algorithm~\ref{alg:trajectory} always terminates in finite time at an equilibrium of the COPS dynamics. The detailed proof is provided in Appendix~\ref{proof:thm:convergence}.

\section{Numerical and Empirical Studies}
\subsection{Survival likelihood and power evolution}
The static power-allocation game may admit multiple Nash equilibria. We employ asynchronous best-response dynamics to enumerate the set of Nash equilibria. In each simulation, given a signed network and the countries' powers, we randomly generate an initial strategy matrix $X(0)$. At each time step, a country $i$ is randomly activated and updates its strategy by computing a best response to the utility function
\begin{align*}
u_i(X)=\!\sum_{j\in \A_i} \mathbb{1}_{\{s_j(X)\le 0\}} 
+ \!\!\!\!\sum_{j\in \F_i\setminus\{i\}} \!\!\mathbb{1}_{\{s_j(X)\ge 0\}} 
+ n\mathbb{1}_{\{s_i(X)\ge 0\}} .
\end{align*}
If no country can strictly increase its utility over 10,000 consecutive updates, the process is declared to have reached a Nash equilibrium. We repeat this procedure for 10,000 independent runs. For each country $i$, we numerically define its \emph{survival likelihood} as the fraction of equilibria in which country $i$ is not in danger.

\begin{figure}[htb]
    \centering
    \includegraphics[width=0.96\linewidth]{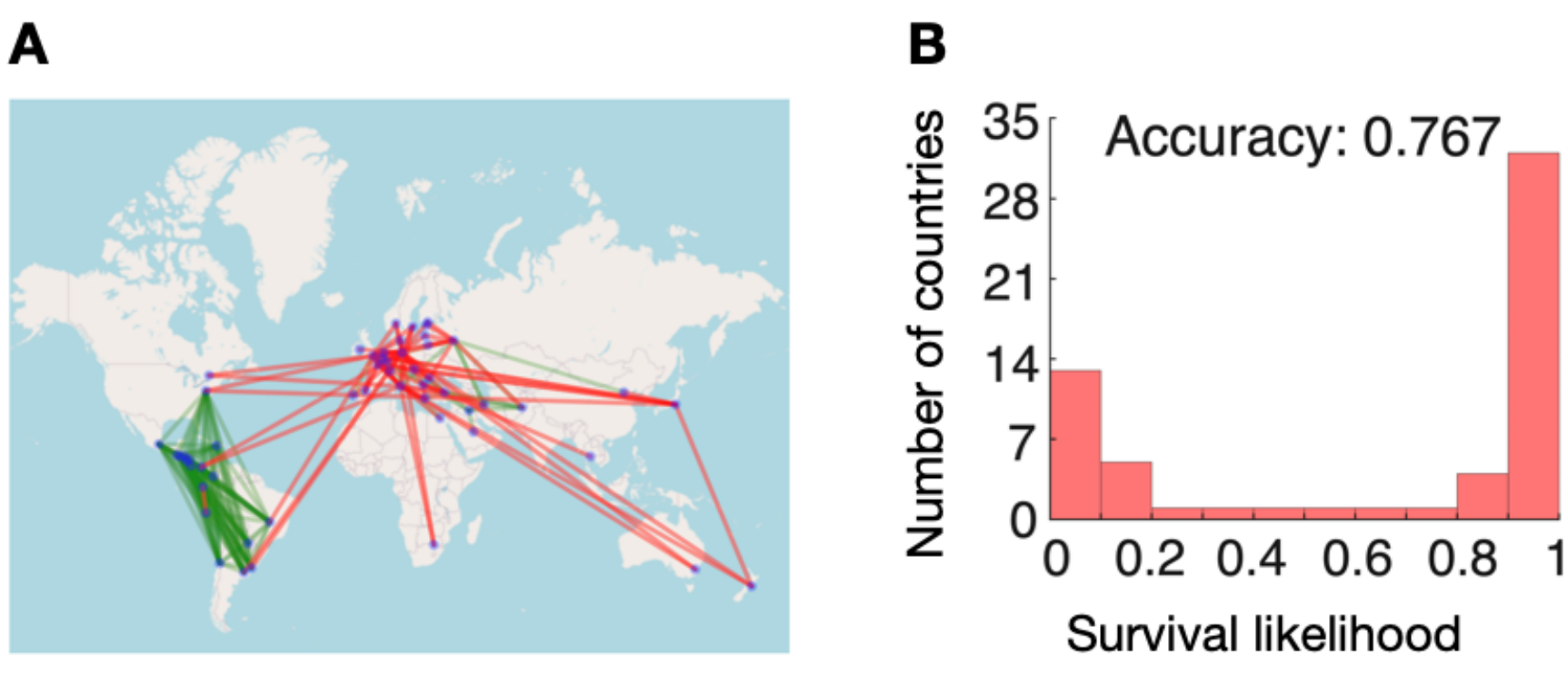}
    \caption{Computation of countries' survival likelihoods based on real data. Panel~\textbf{A} shows the signed network in 1940 constructed via the Correlates of War dataset. Panel~\textbf{B} shows the distribution of countries' survival likelihoods in 1940 computed based on the SPA game model.}
    \label{fig:wide}
\end{figure}

\emph{World War II case study:}
Signed networks are constructed from the Formal Alliances (FA) dataset (version 4.1)~\cite{gibler2008international} and the Dyadic Militarized Interstate Disputes (Dyadic MID) dataset (version 4.02)~\cite{maoz2019dyadic}, where edge signs are determined by whether cooperative incidents outnumber conflict incidents between two countries in a given year (see Fig.~\ref{fig:wide}\textbf{A} for 1940). Countries' powers are taken from the Composite Index of National Capability (CINC) in National Material Capabilities (NMC) dataset (version 6.0)~\cite{singer1988reconstructing}. The FA dataset (version 4.1), Dyadic MID dataset (version 4.02), and NMC dataset (version 6.0) come from the Correlates of War project~\cite{COWdatasets2026}. For each year from 1939 to 1945, we compute each country’s survival likelihood and predict whether a war occurs within its territory in the following year using a threshold of $0.5$.

\zcz{For illustration, Appendix~\ref{apx:WW2} reports the annual relationships among eight countries---China, France, Germany, Italy, Japan, the Soviet Union, the United Kingdom, and the United States---from 1939 to 1945. If one of these countries does not appear in a particular annual table, no data are available for that country in that year. The appendix also reports the corresponding historical states and computed survival likelihoods for all countries included in each annual simulation.}

The survival likelihoods exhibit a highly polarized distribution concentrated near $0$ and $1$ (see Fig.~\ref{fig:wide}\textbf{B}), rendering the predictions robust to the threshold choice. The prediction accuracy exceeds $74\%$ in every year, with an average of $76.8\%$ over the entire period. Given that existing quantitative and network-based studies typically report predictive performance in the range of 60\%--75\%~\cite{MDW-BDG-KMB:10}, the accuracy achieved here is competitive, especially as it is derived from equilibrium outcomes of a structural game model rather than from reduced-form statistical fitting or supervised learning.

\emph{Post-war power dynamics:}
We further construct signed networks from the FA dataset (version 4.1) and Dyadic MID dataset (version 4.02) over the period 1955--2002 and obtain GDP data from the Penn World Table (version 10.0)~\cite{feenstra2015next}, which we use as an empirical proxy for national power. \zcz{As illustrative examples, Appendix~\ref{apx:PW} reports the intercountry relationships among eight countries---Canada, China, France, Italy, Japan, the Soviet Union, the United Kingdom, and the United States---in four selected years---1955, 1967, 1979, and 1991. For the same four years, the appendix also reports the simulated survival likelihoods and indicators of subsequent empirical power growth for all countries included in the corresponding annual simulations.} Using the same simulation procedure, we compute each country's annual survival likelihood and predict whether its national power increases in the subsequent year according to whether the survival likelihood exceeds $0.5$. Comparison with empirical data shows an average prediction accuracy of $80.79\%$ over this period, \zcz{with the annual prediction accuracies reported in Table~\ref{tab:annual_accuracy} of Appendix~\ref{apx:PW}.} These results provide empirical support for the core assumption of the COPS dynamics in Section~III that a country’s power tends to increase when it is not in danger.

\subsection{Power distribution and network structure}

We numerically investigate how the structure of signed networks affects global prosperity and equality under a fixed level of conflict. Specifically, we simulate the COPS dynamics on networks with the same numbers of positive and negative edges but different signed configurations, and examine how the countries' average power and the Gini coefficient of the power distribution depend on the network structure at steady state. We focus in particular on \emph{frustration}, a metric quantifying the deviation of a signed network from structural balance~\cite{facchetti2011computing}. A structurally balanced network admits a partition into two factions with positive intra-faction edges and negative inter-faction edges. Computing frustration is NP-hard; we adopt an efficient greedy heuristic that yields locally optimal solutions.

\begin{figure*}[!t]
    \centering
    \includegraphics[width=\textwidth]{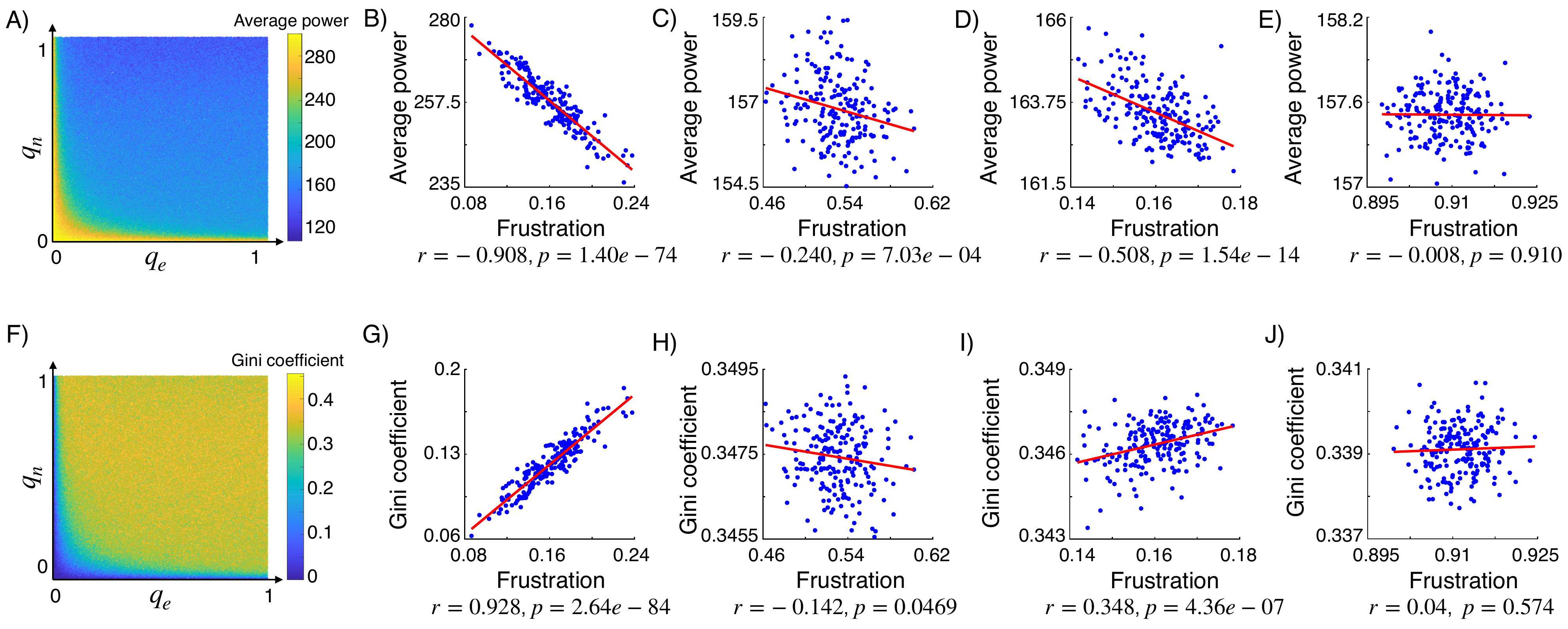}
    \caption{How average power and Gini coefficient at steady state change with edge density, ratio of negative edges, and frustration. Panels~\textbf{A} and~\textbf{F} show heat maps of the average power and the Gini coefficient, respectively, as functions of edge density $q_e$ and ratio of negative edges $q_n$. Panels~\textbf{B}--\textbf{E} show the correlation between average power and frustration for $(q_e,q_n)=(0.08,0.08)$, $(0.08,0.92)$, $(0.92,0.08)$, and $(0.92,0.92)$, respectively. Panels~\textbf{G}--\textbf{J} show the corresponding correlations between the Gini coefficient and frustration. In these panels, $r$ means the Pearson correlation coefficient and $p$ is the p-value. Each data point is one independent simulation. }
    \label{fig:power_gini}
\end{figure*}

The COPS dynamics are simulated on Erd\H{o}s--R\'enyi networks with $100$ nodes, varying edge densities and ratios of negative edges. In each simulation, we generate a signed network with prescribed parameters and compute its frustration, then randomly initialize countries' powers and run the COPS dynamics until convergence. At steady state, we record the average and the Gini coefficient of countries' powers. As shown in Figs.~\ref{fig:power_gini}\textbf{A} and~\ref{fig:power_gini}\textbf{F}, increasing the edge density or the ratio of negative edges reduces the average power while increasing inequality. Moreover, Figs.~\ref{fig:power_gini}\textbf{B}--~\ref{fig:power_gini}\textbf{D} and~\ref{fig:power_gini}\textbf{G}--~\ref{fig:power_gini}\textbf{I} reveal a clear dependence on frustration. In particular, for sparse networks with a low ratio of negative edges (Fig.~\ref{fig:power_gini}\textbf{B} and~\ref{fig:power_gini}\textbf{G}), lower frustration is strongly associated with higher average power and lower inequality, indicating that, under the same level of conflict, more structurally balanced international relations promote a more prosperous and equitable world.

\subsection{Robustness under alternative modeling specifications}

\zcz{The formulation of the COPS dynamics introduced in Section~III assumes that countries' powers and power allocations are nonnegative integers, that powers evolve according to a unit-step rule, and that countries revise their strategies according to Rules~1--4. To examine whether the numerical findings depend critically on these specifications, we consider a combined alternative formulation that simultaneously modifies all three assumptions.}

\begin{figure*}[!t]
    \centering
    \includegraphics[width=\textwidth]{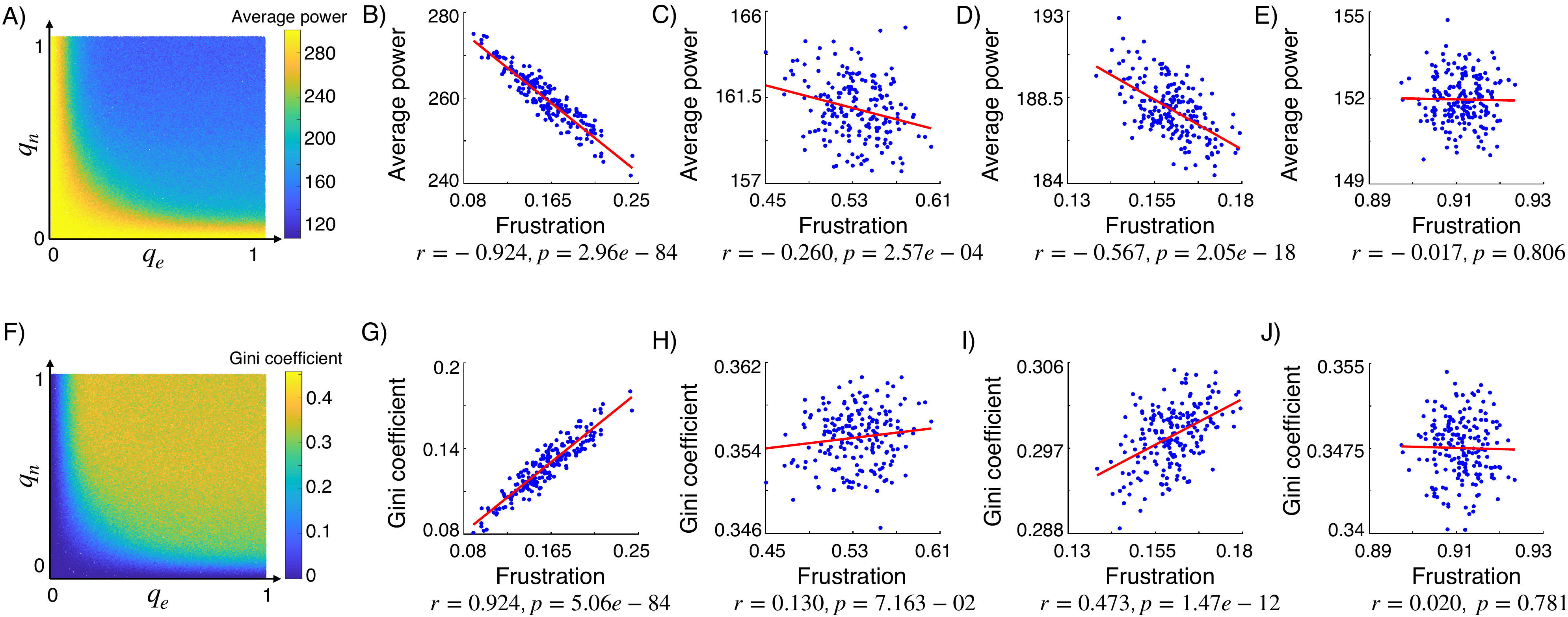}
    \caption{Average power and the Gini coefficient at steady state under the combined alternative formulation. Panels~\textbf{A} and~\textbf{F} show heat maps of the average power and the Gini coefficient, respectively, as functions of edge density $q_e$ and ratio of negative edges $q_n$. Panels~\textbf{B}--\textbf{E} show the correlations between average power and frustration for $(q_e,q_n)=(0.08,0.08)$, $(0.08,0.92)$, $(0.92,0.08)$, and $(0.92,0.92)$, respectively. Panels~\textbf{G}--\textbf{J} show the corresponding correlations between the Gini coefficient and frustration. In these panels, $r$ means the Pearson correlation coefficient and $p$ denotes the p-value. Each data point represents one independent simulation.}
    \label{fig:power_gini_real}
\end{figure*}

\zcz{First, countries' powers and power allocations are allowed to take nonnegative real values. Second, the unit-step power update is replaced by a stochastic state-dependent mechanism, under which the change in an activated country's power is randomly generated from a distribution whose expected value is monotonically increasing in the country's state. Third, after its power is updated, the activated country computes and adopts a best-response strategy rather than updating its strategy according to Rules~1--4. All other network-generation procedures and simulation settings are kept the same as in the original simulations. The corresponding results are shown in Fig.~\ref{fig:power_gini_real}.}

\zcz{Under this combined alternative formulation, all tested trajectories reach steady states within the prescribed simulation horizon. The principal qualitative patterns obtained under the original COPS dynamics are largely preserved. In particular, increasing the edge density or the ratio of negative edges generally reduces the average power and increases inequality. Moreover, for sparse networks with a low ratio of negative edges, lower frustration remains strongly associated with higher average power and a lower Gini coefficient.}

\zcz{One difference arises for sparse networks with a high ratio of negative edges, i.e., $(q_e,q_n)=(0.08,0.92)$. Under the original formulation, frustration and the Gini coefficient exhibit a weak and marginally significant negative correlation (Fig.~\ref{fig:power_gini}\textbf{H}; $r=-0.142$, $p=0.0469$), whereas under the alternative formulation, the correlation becomes weakly positive and is not statistically significant at the $5\%$ level (Fig.~\ref{fig:power_gini_real}\textbf{H}; $r=0.130$, $p=0.0716$). Since both correlations are weak, this difference does not affect the main qualitative conclusions regarding average power, inequality, and structural balance. Overall, the results indicate that the numerical findings are robust to simultaneously allowing real-valued powers and allocations, adopting a smoother stochastic mechanism for power evolution, and replacing the original strategy-update rules with best-response updates. Since the real-valued formulation has an infinite state space, these simulations provide numerical evidence of robustness rather than an extension of the finite-time convergence result established in Theorem~\ref{thm:convergence}.}

\section{Conclusion and Further Discussion}

This paper studies power allocation games on signed networks. We modified an earlier model~\cite{li2022games} by Li and Morse so that the players' behavior is more strategic. For the static power-allocation game, we prove the existence of pure Nash equilibria and analyze the equilibria in special network configurations. Then we extend the model to a dynamic setting and prove its convergence to the Nash equilibria of the static model. Using real-world data, we validate the proposed framework through numerical simulations. The results show that the equilibrium-based survival likelihood provides effective predictions of both conflict occurrence and national power evolution. Moreover, simulations of the dynamic model reveal that, under a fixed level of conflict, signed networks that are closer to structural balance tend to promote higher average power and lower inequality across countries. More broadly, our framework opens the door to integrating signed network structure, strategic interaction, and endogenous power evolution within a unified game-theoretic model of international relations. Future work includes extending the model to scenarios of weighted signed networks or partial-information games, and developing inverse methods to infer inter-country relationship weights from observed equilibrium power distributions.

\appendices

\section{Proof of Lemma~\ref{lem:impossible-all-danger}}\label{proof:lem:impossible-all-danger}

By definition, 
    \begin{align*}
        \sum_{i\in \V}s_i(X) = \sum_{k\in \V}\sum_{l\in \F_k} x_{lk} + \sum_{i\in \V} \sum_{j\in \A_i} x_{ij} -\sum_{r\in \V}\sum_{s\in \A_r} x_{sr}.
    \end{align*}
For any $i\in \V$ and any $j\in \A_i$, since $i\in \A_j$ also holds, the term $x_{ij}$ in the summation $\sum_{i\in \V} \sum_{j\in \A_i} x_{ij}$ is also in the summation $\sum_{r\in \V}\sum_{s\in \A_r} x_{sr}$, with $r=j$ and $s=i$. Following the same argument, any term in the summation $\sum_{r\in \V}\sum_{s\in \A_r} x_{sr}$ is also included in $\sum_{i\in \V} \sum_{j\in \A_i} x_{ij}$. Therefore, we have 
\begin{align*}
    \sum_{i\in \V} \sum_{j\in \A_i} x_{ij} -\sum_{r\in \V}\sum_{s\in \A_r} x_{sr}=0, 
\end{align*}
and thereby
\begin{align*}
    \sum_{i\in \V}s_i(X) = \sum_{k\in \V}\sum_{l\in \F_k} x_{lk}\ge 0.
\end{align*}
If every country is in danger under the strategy matrix $X$, then $s_i(X)<0$ for any $i\in \V$ and thus $\sum_{i\in \V}s_i(X)<0$, which leads to a contradiction. Therefore, at least one country is either safe or precarious.

Now consider the case in which there exists at least one country in danger under $X$. Define
\begin{align*}
\V^-\!=\{k\in \V\,|\, s_k(X)<0\} \,\text{ and }\, \V^+\!=\{k\in \V\,|\, s_k(X)\ge 0\}.
\end{align*}
Suppose $x_{ij}=0$ for any $i\in \V$ and any $j\in \A_i$ such that $s_i(X)\ge 0$ and $s_j(X)<0$. We will show that this assumption leads to a contradiction. By definition, we have
\begin{align*}
    \sum_{k\in \V^-} s_k(X) & = \sum_{k\in \V^-}\Big( \sum_{r\in \F_k} x_{rk} +\sum_{r\in \A_k} x_{kr} - \sum_{r\in \A_k} x_{rk} \Big)\\
    & = \sum_{k\in \V^-} \sum_{r\in \F_k} x_{rk} +\sum_{k\in \V^-}\sum_{r\in \A_k\cap \V^-} x_{kr}\\
    &\quad +\sum_{k\in \V^-}\sum_{r\in \A_k\cap \V^+}x_{kr} - \sum_{k\in \V^-}\sum_{r\in \A_k\cap \V^-} x_{rk}.
\end{align*}
Since, for any $i,j\in \V$, $j\in \A_i$ if and only if $i\in \A_j$, any term in the summation $\sum_{k\in \V^-}\sum_{r\in \A_k\cap \V^-}x_{kr}$ will also be included in the summation $\sum_{k\in \V^-}\sum_{r\in \A_k \cap \V^-}x_{rk}$, and vice versa. Therefore, 
\begin{align*}
    \sum_{k\in \V^-}\sum_{r\in \A_k\cap \V^-}x_{kr} - \sum_{k\in \V^-}\sum_{r\in \A_k \cap \V^-}x_{rk} = 0.
\end{align*}
As a result, $\sum_{k\in \V^-}s_k(X)\ge \sum_{k\in \V^-}\sum_{r\in \F_k} x_{rk}\ge 0$, which contradicts the fact that $s_k(X)<0$ for any $k\in \V^-$. This contradiction completes the proof.\hfill$\blacksquare$

\section{Proof of Lemma~\ref{lem:precarious number}}\label{proof:lem:precarious number}

Given any $X\in \OOmega$, any $i\in \V$, and any $X^+\in \bfU_i(X)$, let
\begin{equation}
\label{eq:def}
\begin{aligned}
    \C_i^+(X^+,X) &= \Big{\{} j\neq i\,\Big|\, u_{ji}(X)=0\text{ and }u_{ji}(X^+)=1 \Big{\}}\\
    \C_i^-(X^+,X) &= \Big{\{} j\neq i\,\Big|\, u_{ji}(X)=1\text{ and }u_{ji}(X^+)=0 \Big{\}}\\
    \C_i^0(X^+,X) &= \Big{\{} j\neq i\,\Big|\, u_{ji}(X)=u_{ji}(X^+) \Big{\}}.
\end{aligned}
\end{equation}
We use $\C_i^+$, $\C_i^-$, and $\C_i^0$ as shorthand for the three sets above, respectively. These sets and $\{i\}$ form a partition of $\V$.

For any $j\in \C_i^+$, $u_{ji}(X^+)=1$ implies that $j\in \A_i\cup \F_i$, and $u_{ji}(X)=0$ implies that country $j$ is non-precarious under $X$. Moreover, Rule 1 for $\bfU_i(X)$ requires that
\begin{align*}
    x^+_{ij}-x_{ij}\le |s_j(X)|.
\end{align*}
Therefore, from $X$ to $X^+$, $j$'s safety state cannot be changed from ``safe'' to ``in danger'', or vice versa. Moreover, country $j$ does contribute one unit of utility to $i$ under $X^+$. Hence country $j$ must be precarious under $X^+$. Consequently,
\begin{align*}
    N_p\big( \C_i^+,X^+ \big) - N_p\big( \C_i^+,X \big)=\card\big( \C_i^+ \big).
\end{align*}

For any $j\in \C_i^-$, $N_p(\C_i^-,X^+)-N_p(\C_i^-,X)\ge -\card(\C_i^-)$ holds by definition.

Now consider any $j\in \C_i^0$. If $j\notin \F_i\cup \A_i$, then $i$'s power reallocation does not affect $j$'s safety state. Therefore, $s_j(X^+)=s_j(X)$. If $j\in \F_i$, then $j\in \C_i^0$ and Rule~1 (no unnecessary increase in power allocated to other countries) requires that either $\sgn(s_j(X))=\sgn(s_j(X^+))$ or $s_j(X)>s_j(X^+)=0$. In this case, we have $\mathbb{1}_{\{s_j(X^+)=0\}}\ge \mathbb{1}_{\{s_j(X)=0\}}$. If $j\in \A_i$, then $j\in \C_i^0$ and Rule~1 requires that either $\sgn(s_j(X))=\sgn(s_j(X^+))$ or $s_j(X)<s_j(X^+)=0$, which also leads to $\mathbb{1}_{\{s_j(X^+)=0\}}\ge \mathbb{1}_{\{s_j(X)=0\}}$. Therefore, 
\begin{align*}
    N_p&(\C_i^0,X^+)-N_p(\C_i^0,X)\\
    &=\sum_{j\in \C_i^0}\Big( \mathbb{1}_{\{s_j(X^+)=0\}}- \mathbb{1}_{\{s_j(X)=0\}} \Big)\ge 0.
\end{align*}

According to the update rule for countries' powers, $s_i(X)\ge 0$ implies that $(X^+\mathbf{1}_n)_i\ge (X\mathbf{1}_n)_i$. In this case, according to Lemma~\ref{lem:safe-prec-utility-nondecrease}, the candidate set $\bfU_i(X)$ is either $\OOmega_2$ or $\OOmega_3$, and $s_i(X^+)\ge 0$. Therefore, 
\begin{align*}
    u_i(X^+)-u_i(X) &= \sum_{j\neq i} u_{ji}(X^+)-\sum_{j\neq i} u_{ji}(X)\\
    &=\card(\C_i^+)-\card(\C_i^-).
\end{align*}

Combining the sets $\C_i^+$, $\C_i^-$ and $\C_i^0$, we have
\begin{equation}
\label{eq:prenum}
\begin{aligned}
    N_p&(\V\setminus\! \{i\},X^+)-N_p(\V\setminus\! \{i\},X) \\
    &=N_p(\C_i^+,X^+)-N_p(\C_i^+,X) + N_p(\C_i^-,X^+)\\
    &\quad -N_p(\C_i^-,X) + N_p(\C_i^0,X^+)-N_p(\C_i^0,X)\\
    &\ge u_i(X^+)-u_i(X)+ N_p(\C_i^0,X^+)-N_p(\C_i^0,X)\\
    &\ge u_i(X^+)-u_i(X).
\end{aligned}
\end{equation}

If $s_i(X)>0$, then 
\begin{align*}
    N_p&(\V,X^+)-N_p(\V,X)\\
    &= N_p(\V\setminus \!\{i\},X^+)-N_p(\V\setminus\! \{i\},X) \\
    &\quad+\mathbb{1}_{\{s_i(X^+)=0\}}-\mathbb{1}_{\{s_i(X)=0\}}\\
    &\ge u_i(X^+)-u_i(X) +\mathbb{1}_{\{s_i(X^+)=0\}}-0\ge 0;
\end{align*}
If $s_i(X)=0$ and $\bfU_i(X)=\OOmega_2$, then $u_i(X^+)-u_i(X)\ge 1$. Therefore,
\begin{align*}
    N_p&(\V,X^+)\!-\!N_p(\V,X)\\
    &\ge \mathbb{1}_{\{s_i(X^+)=0\}}-\mathbb{1}_{\{s_i(X)=0\}} + u_i(X^+)-u_i(X)\\
    &\ge -1+1=0;
\end{align*}
If $s_i(X)=0$ and $\bfU_i(X)=\OOmega_3$, then $u_i(X^+)=u_i(X)$. Moreover, according to Rule 3, country $i$ can only retract some power spent on attacking its enemies and allocate the retracted power to itself. According to Definition~\ref{def:state}, doing this will not change $i$'s safety state, i.e., $s_i(X^+)=s_i(X)=0$. Therefore, 
\begin{align*}
    N_p&(\V,X^+)-N_p(\V,X)\\
    &\ge\mathbb{1}_{\{s_i(X^+)=0\}}-\mathbb{1}_{\{s_i(X)=0\}} +u_i(X^+)-u_i(X)=0.
\end{align*}
This concludes the proof that $N_p(\V,X^+)\ge N_p(\V,X)$.\hfill$\blacksquare$

\section{Proof of Lemma~\ref{lem:S1}}\label{proof:lem:S1}
According to Lemma~\ref{lem:precarious number}, strategy updates by safe or precarious countries preserve or increase the total number of precarious countries, which is upper bounded by $n$. Consequently, for any $X(0)\in\OOmega$, there exists a feasible trajectory $\{X(t)\}$ and a finite time $T_1$ such that $X(T_1)$ belongs to $\bfS_1$. 

For any $X\in \bfS_1$, if a safe country $i$ updates its strategy, then we have $\mathbb{1}_{\{s_i(X^+)=0\}}-\mathbb{1}_{\{s_i(X)=0\}}\ge 0$, for any $X^+\in \bfU_i(X)$ and Lemma~\ref{lem:safe-prec-utility-nondecrease} implies that $s_i(X^+)\ge 0$. We claim that no $X^+\in \bfU_i(X)$ can satisfy $u_i(X^+)\ge u_i(X)+1$. Otherwise, Lemma~\ref{lem:precarious number} yields
\begin{align*}
    N_p(\V\setminus \{i\},X^+)-N_p(\V\setminus \{i\},X)
    \ge u_i(X^+)-u_i(X)\ge 1,
\end{align*}
and thus $N_p(\V,X^+)-N_p(\V,X)\ge 1$, contradicting the definition of $\bfS_1$. Therefore, no candidate strategy can strictly increase $i$'s utility. By Lemma~\ref{lem:safe-prec-utility-nondecrease}, this implies $\bfU_i(X)=\OOmega_3$, i.e., $i$ can only retract power spent on attacking its enemies, and reallocate the retracted power and the increased power (if any) to itself; such an update preserves its safety state.

Moreover, since the utility of country $i$ remains unchanged, its new strategy does not change the safety state of any enemy from ``in danger'' or ``precarious'' to ``safe''; consequently, $i$ cannot change the safety state of any enemy from ``in danger'' to ``precarious'' either, since this would alter the total number of precarious countries. This concludes the proof.\hfill$\blacksquare$

\section{Proof of Lemma~\ref{lem:S2}}\label{proof:lem:S2}
For any $X(0)\in\bfS_1$, we activate only precarious countries. According to the definition of $\bfS_1$, $X(t)$ will still be in $\bfS_1$ for any $t\ge 0$.

We first examine the consequence of a strategy update by a precarious country that strictly increases its utility. Suppose $u_{i_t}(X(t))\ge u_{i_t}(X(t-1))+1$ at some $t\ge 1$. By Lemma~\ref{lem:precarious number}, 
\begin{align*}
    N_p(\V\setminus \{i_t\},X(t))-N_p(\V\setminus\{i_t\},X(t-1))\ge 1.
\end{align*}
In addition, $s_{i_t}(X(t-1))=0$ implies that
\begin{align*}
    \mathbb{1}_{\{s_{i_t}(X(t))=0\}}-\mathbb{1}_{\{s_{i_t}(X(t-1))=0\}}\ge -1.
\end{align*}
Since $X(t-1)\in \bfS_1$, the total number of precarious countries must remain unchanged. Therefore, the two inequalities above hold with equality, which implies $s_{i_t}(X(t))\ge 1$, and
\begin{align*}
    N_p\big(\V\setminus \{i_t\},X(t)\big)&=N_p\big(\V\setminus \{i_t\},X(t-1)\big)+1.
\end{align*}

For any $X\in \OOmega$, define $\Aid(X)=\sum_{i\in \V}\sum_{j\in \F_{i}\setminus\{i\}}x_{ij}$, i.e., the total support between friendly countries. Note that
\begin{align*}
    s_{i_t}\big( X(t) \big) &= \big( X(t)\mathbf{1}_n \big)_{i_t} - \!\!\sum_{j\in \F_{i_t}\setminus\{i_t\}}x_{i_tj}(t)\\
    &\quad+\!\!\sum_{j\in \F_{i_t}\setminus\{i_t\}}x_{ji_t}(t) - \!\!\sum_{j\in \A_{i_t}}x_{ji_t}(t),\\
    x_{kj}(t) & =x_{kj}(t-1),\quad \forall\, k\neq i_t\text{ and }\forall\, j\in \V, \text {and}\\
    \big( X(t)\mathbf{1}_n \big)_{i_t} & = \big( X(t-1)\mathbf{1}_n \big)_{i_t},\quad \text{since }s_{i_t}\big( X(t-1) \big)=0.
\end{align*}
We thereby obtain
\begin{align*}
    \Aid&(X(t))-\Aid(X(t-1))\\
    &=\sum_{i\in\V}\sum_{j\in \F_{i}\setminus \{i\}} x_{i j}(t)-\sum_{i\in\V}\sum_{j\in \F_{i}\setminus \{i\}} x_{i j}(t-1)\\
    &=\sum_{j\in \F_{i_t}\setminus \{i_t\}} x_{i_t j}(t)-\sum_{j\in \F_{i_t}\setminus \{i_t\}} x_{i_t j}(t-1)\\
    & = s_{i_t}\big( X(t-1) \big)-s_{i_t}\big( X(t) \big)\leq -1.
\end{align*}
Hence, whenever a precarious country strictly increases its utility via a strategy update, the total friendly support $\Aid(X)$ decreases by at least one. Since $\Aid(X(t))$ is nonnegative, such utility-increasing updates can occur only finitely many times. Therefore, there exists a finite time $T_2\ge 0$ such that $X(T_2)$ enters the following subset of $\bfS_1$:
\begin{align*}
    \bfS_2'=\Big{\{} & X\in \bfS_1\,\Big|\, u_{i_t}\big( \tilde{X}(t) \big)=u_{i_t}\big( \tilde{X}(t-1) \big),\,\, \forall\, t\in \mathbb{N}_+,\text{ for}\\
    &\text{any }\{i_t\} \text{ and any }\{\tilde{X}(t)\} \text{ such that } \tilde{X}(0)=X\\
    & \text{and }s_{i_t}\big( \tilde{X}(t-1) \big)=0,\,\, \forall\, t\in \mathbb{N}_+.\Big{\}}
\end{align*}

We next prove that $\bfS_2'=\bfS_2$. Fix any $X\in \bfS_2'$ and suppose that a precarious country $i$ is activated. Since $u_i(X^+)=u_i(X)$ for any $X^+\in \bfU_i(X)$, Lemma~\ref{lem:safe-prec-utility-nondecrease} implies that $\bfU_i(X)=\OOmega_3$. Hence, $i$ can only retract power spent on attacking its enemies and reallocate it to itself, without changing its own safety state or the safety states of its friends.

For any $j\in \A_i$, since $x_{ij}^+\le x_{ij}$, we have $s_j(X^+)\ge s_j(X)$ and thus $u_{ji}(X^+)\le u_{ji}(X)$. Because $u_i(X^+)=u_i(X)$, there is no $j\in \A_i$ such that $s_j(X^+)>0$ and $s_j(X)\leq 0$. Moreover, since $X\in \bfS_1$ and $s_i(X)=0$, the total number of precarious countries must remain unchanged, which excludes the possibility that $s_j(X^+)=0$ and $s_j(X)<0$ for any $j\in \A_i$. Therefore, $\sgn\big( s_j(X^+) \big)=\sgn\big( s_j(X) \big)$ for any $j\in \A_i$. That is, no country's safety state changes, and hence $\bfS_2'\subseteq \bfS_2$.

Conversely, $X\in \bfS_2$ immediately implies $u_i(X^+)=u_i(X)$ for the activated country $i$. Therefore, $\bfS_2'=\bfS_2$. \hfill$\blacksquare$

\section{Proof of Lemma~\ref{lem:no-precarious-initially}}\label{proof:lem:no-precarious-initially}

Define 
\begin{align*}
    S^*=\Big{\{} X\in \OOmega \,\Big|\,& N_p(\V,X)>0\\
    & \text{or }\bfU_i(X)=\{X\}\text{ for any }i\in \V\Big{\}}.
\end{align*}
That is, a strategy matrix in $S^*$ either contains at least one precarious country or is an equilibrium of the COPS dynamics. Let $X(0)\in\OOmega$ be an arbitrary initial state satisfying $N_p(\V,X(0))=0$. We construct a feasible trajectory as follows and then show that it reaches $S^*$ in finite time. At each time $t\in \mathbb{N}_+$, execute one of the following procedures until $N_p(\V,X(t))>0$.

\emph{Procedure 1:} If there exists a safe country $i\in \V$ such that $u_i(X^+)>u_i(X(t-1))$ for some $X^+\in \bfU_i(X(t-1))$, randomly choose such a country $i$ and update its power allocation accordingly.

\emph{Procedure 2:} Otherwise, randomly choose a safe country $i$ and update its power allocation by retracting one unit of power spent on attacking one of $i$'s enemies (if any) and allocating the retracted power, together with the increased power (if any), to $i$ itself. By Lemma~\ref{lem:impossible-all-danger}, a safe country must exist whenever $N_p(\V,X(t-1))=0$. Moreover, since no safe country can strictly increase its utility via power reallocation, Lemma~\ref{lem:safe-prec-utility-nondecrease} implies that the selected country updates its strategy according to Rule~3. Hence, Procedure~2 is always feasible.

If Procedure~1 occurs at some finite time $T\in \mathbb{N}_+$, then
\begin{align*}
    N_p&(\V,X(T)) -  N_p(\V,X(T-1))\\
    & \ge \mathbb{1}_{\{ s_{i_T}(X(T))=0 \}} - \mathbb{1}_{\{ s_{i_T}(X(T-1))=0 \}}\\
    &\quad + u_{i_T}(X(T)) - u_{i_T}(X(T-1))>0,
\end{align*}
which implies that $X(T)\in S^*$.

Suppose that Procedure~1 never occurs and Procedure~2 is executed repeatedly. Since the total power allocated between antagonistic countries is upper bounded by $nK$ and each country's power is upper bounded by $K$, $\{X(t)\}$ must enter one of the following two scenarios at some finite time $T$:
\begin{enumerate}
    \item \emph{Case 1:} Every country is safe and its power reaches the upper bound $K$, and no country allocates power to attack any other country. In this case, since no country can strictly increase its utility via power reallocation, Rule~3 implies that $\bfU_i(X(T))=\{X(T)\}$ for all $i\in \V$. Hence, $X(T)$ is an equilibrium, and thus $X(T)\in S^*$.
    \item \emph{Case 2:} There exists at least one country in danger, but no safe country allocates power to attack any country in danger. By Lemma~\ref{lem:impossible-all-danger}, such a situation is incompatible with the absence of precarious countries. Therefore, $X(T)$ must contain at least one precarious country, and hence $X(T)\in S^*$.
\end{enumerate}
Therefore, for any given $X(0)\in \OOmega$ with $N_p(\V,X(0))=0$, there always exists a feasible trajectory that reaches $S^*$ in finite time. This concludes the proof.\hfill$\blacksquare$

\section{Proof of Theorem~\ref{thm:convergence}}\label{proof:thm:convergence}
We first show that the four phases in Algorithm 1 are feasible and terminate in finite time. By Lemmas~\ref{lem:S1} and~\ref{lem:S2}, the first two phases are feasible and terminate in finite time. Hence, the hitting times $T_1$ and $T_2$ are both finite.

Since $\bfS_2$ is invariant with respect to strategy updates by precarious countries, the trajectory remains in $\bfS_2$ throughout Phase 3. Moreover, by the definition of $\bfS_2$, such updates do not change any country's safety state, and therefore the updated strategy matrix at each time step is chosen from $\OOmega_3$. Therefore, Phase 3 is feasible. Since the total amount of power allocated to attacking non-precarious countries is finite, Phase 3 must terminate at some finite time $T_3\ge T_2$, when no precarious country allocates power to attack any non-precarious enemy.

During Phase 4, the trajectory $X(t)$ may no longer remain in $\bfS_2$, but, by definition of $\bfS_1$, it always remains in $\bfS_1$. Hence, by Lemma~\ref{lem:S1}, no country's safety state changes throughout Phase 4, and every updated strategy matrix is chosen from $\OOmega_3$. Therefore, Phase 4 is also feasible. By the same argument as in Phase 3, Phase 4 must terminate at some finite time $T_4\ge T_3$, when every safe country's power has reached the upper limit $K$ and no safe country allocates power to attack any non-precarious enemy.

According to Lemma~\ref{lem:impossible-all-danger}, if there exists at least one country in danger, then at least one such country must be attacked by a safe or precarious country. Therefore, the only possibility at time $T_4$ is that no country is in danger. Moreover, since no country's safety state changes along the trajectory $\{X(t)\}$ from $T_2$ to $T_4$, there is no country in danger at time $T_2$ either.

The preceding arguments establish the following properties
of the trajectory and its terminal strategy matrix:
\begin{enumerate}[(P1)]
\renewcommand{\theenumi}{P\arabic{enumi}}
\renewcommand{\labelenumi}{(\theenumi)}

    \item\label{prop:belong_S1}
    For every $t\in\{T_2,\dots,T_4\}$, $X(t)\in\bfS_1$.

    \item\label{prop:state-unchanged}
    From $T_2$ to $T_4$, every country's safety state remains
    unchanged and no country is in danger under $X(t)$.

    \item\label{prop:no-attack-safe}
    Under $X(T_4)$, every safe country's power equals the
    upper limit $K$, and no safe country is attacked by any
    country.

\end{enumerate}

We next prove that the terminal strategy matrix $X(T_4)$ is an equilibrium of the COPS dynamics. Suppose, to the contrary, that $X(T_4)$ is not an equilibrium. Then there exists $i\in \V$ such that $\bfU_i(X(T_4))\neq \{X(T_4)\}$. By property~(\ref{prop:state-unchanged}), $s_i(X(T_4))\geq0$ for every $i\in\V$. By Lemma~\ref{lem:safe-prec-utility-nondecrease}, regardless of which country $i$ is activated, any feasible update must follow either Rule~2 or Rule~3, while also satisfying Rule~1.

Consider the case in which a country $i$ updates its strategy via Rule~3. In this case, $\bfU_i(X(T_4))=\OOmega_3$, i.e., country $i$ either adds the increased power to itself, or retracts some power spent on attacking its enemies, or does both. However, by property~(\ref{prop:no-attack-safe}), country $i$'s power remains unchanged. Moreover, country $i$ cannot retract power from any enemy: safe enemies are not attacked under $X(T_4)$ by property~(\ref{prop:no-attack-safe}), while retracting power from a precarious enemy would make that enemy safe and thereby decrease country $i$'s utility. Therefore, the only strategy in $\OOmega_3$ is $X(T_4)$ itself.

Now consider the case in which country $i$ updates its strategy via Rule~2. Fix any $X^+(T_4)\in\bfU_i(X(T_4))$. By the definition of $\bfU_i(X(T_4))$, we have $u_i(X^+(T_4))\ge u_i(X(T_4))+1$. In the remainder of the proof, we show that this case also leads to a contradiction. 

By properties~(\ref{prop:belong_S1}) and~(\ref{prop:state-unchanged}), $X(T_4)\in\bfS_1$ and no country is in danger under $X(T_4)$. According to Lemma~\ref{lem:S1}, a strategy update by a safe country does not change any country's safety state and therefore cannot strictly increase the activated country's utility. Hence, country $i$ is precarious under $X(T_4)$. By property~(\ref{prop:state-unchanged}), country $i$ remains precarious from $T_2$ to $T_4$. Consequently, its total available power does not change, and thereby $\sum_{j\in \V}x_{ij}^+(T_4)=\big( X(T_2)\mathbf{1}_n \big)_i$. That is, the allocation $(x_{i1}^+(T_4),\dots,x_{in}^+(T_4))$ satisfies the total-power constraint for country $i$ at time $T_2$. We then investigate the following question: Since country $i$ can strictly increase its own utility at time $T_4+1$ by adopting the power allocation $(x_{i1}^+(T_4),\dots, x_{in}^+(T_4))$, would the same allocation also
strictly increase its utility at time $T_2+1$ if it were activated then? For any $t\in \{T_2,\dots, T_4\}$, define $X^+(t)=(x_{ij}^+(t))_{n\times n}$ as 
\begin{align*}
    x_{ik}^+(t) & = x_{ik}^+(T_4), &&\forall\, k\in \V,\\ 
    x_{jk}^+(t) & = x_{jk}(t), &&\forall\, j\neq i, \,\,\forall\, k\in \V.
\end{align*}

Suppose that $u_i(X^{+}(T_2))>u_i(X(T_2))$. Since $X^+(T_2)$ may not satisfy Rule~1 relative to $X(T_2)$, we construct from it a strategy matrix $\widehat{X}^+(T_2)$ that satisfies Rule~1 without decreasing country $i$'s utility. Define, for every $j\neq i$,
\[
b_{ij}=
\begin{cases}
x_{ij}(T_2), 
    & u_{ji}(X(T_2))=1,\\
x_{ij}(T_2)+|s_j(X(T_2))|, 
    & u_{ji}(X(T_2))=0.
\end{cases}
\]
Set $\widehat{x}_{ij}^{+}(T_2) = \min\left\{x_{ij}^{+}(T_2),b_{ij} \right\}$ for all $j\neq i$, and allocate the remaining power to country $i$ itself. All other rows of $\widehat{X}^{+}(T_2)$ are identical to those of $X(T_2)$. By construction, $\widehat X^+(T_2)$ is obtained from $X^+(T_2)$ by making the smallest componentwise reductions necessary to satisfy Rule~1. More precisely, an allocation $x_{ij}^+(T_2)$ is left unchanged whenever it already satisfies the corresponding upper bound in Rule~1, and is otherwise reduced exactly to that upper bound. This minimal truncation preserves every neighbor-related utility term $u_{ji}$, while the truncated power is reallocated to country $i$ itself. Consequently, $u_i(\widehat X^+(T_2))\geq u_i(X^+(T_2))$.

It follows that $u_i(\widehat{X}^{+}(T_2)) > u_i(X(T_2))$. Therefore, $\widehat{X}^{+}(T_2)\in\OOmega_2=\bfU_i(X(T_2))$. However, since $X(T_2)\in\mathbf{S}_2$ and country $i$ is precarious under $X(T_2)$, every feasible update by country $i$ preserves all countries' safety states and, consequently, preserves country $i$'s utility. This yields a contradiction. Therefore, $u_i(X^{+}(T_2))\leq u_i(X(T_2))$.

We next show that the two inequalities $u_i(X^+(T_4))\ge u_i(X(T_4))+1$ and $u_i(X^+(T_2))\le u_i(X(T_2))$ imply the existence of a time $t^*\in\{T_2,\dots,T_4\}$ at which country $i$ admits a feasible strategy update that increases the number of precarious countries. For any $t\in \{T_2,\dots, T_4\}$, we have
\begin{align*}
    s_j(X^+(t)) & =s_j(X(t)) + x_{ij}(t) - x_{ij}^+(T_4), &&\forall\, j\in \A_i,\\
    s_j(X^+(t)) & = s_j(X(t)) - x_{ij}(t) + x_{ij}^+(T_4), &&\forall\, j\in \F_i\setminus\{i\}. 
\end{align*}
Moreover, by property~(\ref{prop:state-unchanged}), no country's safety state ever changes from $T_2$ to $T_4$. Combining this with the construction of Phase~3 and Phase~4, we obtain that, for any $k\in \V$ and any $t\in \{T_2,\dots, T_4-1\}$,
\begin{equation}\label{eq:conv-T2T4-basic}
    \begin{aligned}
    u_k(X(t+1))& = u_k(X(t)), \quad x_{kk}(t+1) \ge x_{kk}(t),\\
    x_{kj}(t+1)&\le x_{kj}(t),\quad \forall\, j\in \A_k, \\
    x_{kj}(t+1)&=x_{kj}(t),\quad \forall\, j\in \F_k\setminus \{k\},\\
    s_k(X(t+1)) & \ge s_k(X(t)),\quad\text{if }s_k(X(T_2))>0,\\
    s_k(X(t+1)) & = s_k(X(t)),\quad\text{if }s_k(X(T_2))=0,\\
    \big( X(t+1)\mathbf{1}_n \big)_k&=\big( X(t)\mathbf{1}_n \big)_k,\ \text{if }s_k(X(T_2))=0.
    \end{aligned}
\end{equation}
For any $t\in \{T_2,\dots, T_4-1\}$ and any $j\in \A_i$,
\begin{align*}
    &s_j(X(t))+x_{ij}(t)\\
    & = \sum_{k\in \F_j\setminus \{j\}}x_{kj}(t) + x_{jj}(t) + \sum_{k\in \A_j} x_{jk}(t) - \sum_{k\in \A_j\setminus\{i\}}x_{kj}(t),
\end{align*}
where $\sum_{k\in \F_j\setminus \{j\}}x_{kj}(t)$ is constant, $x_{jj}(t)+\sum_{k\in \A_j}x_{jk}(t)$ is non-decreasing, and $\sum_{k\in \A_j\setminus\{i\}} x_{kj}(t)$ is non-increasing, due to the definition of $\OOmega_3$. Consequently,
\begin{equation}\label{eq:conv-s+-increasing}
    s_j(X(t+1))+x_{ij}(t+1)\ge s_j(X(t))+x_{ij}(t),
\end{equation}
for any $t\in \{T_2,\dots, T_4-1\}$ and any $ j\in \A_i$, and hence $s_j(X^+(t))\le s_j(X^+(t+1))$, which in turn implies $u_{ji}(X^+(t))\ge u_{ji}(X^+(t+1))$. Therefore, $\sum_{j\in \A_i} u_{ji}(X^+(T_2))\ge \sum_{j\in \A_i} u_{ji}(X^+(T_4))$. Then the inequality $u_i(X^+(T_4))\ge u_i(X(T_4))+1 = u_i(X(T_2))+1$
together with inequalities~\eqref{eq:conv-T2T4-basic} and~\eqref{eq:conv-s+-increasing}, implies that
\begin{align*}
    \sum_{j\in \F_i}  u_{ji}(X^+(T_4)) \ge \sum_{j\in\F_i} u_{ji}(X^+(T_2)) +\Delta u +1,
\end{align*}
where $\Delta u = u_i(X(T_2))-u_i(X^+(T_2))\ge 0$. For any $j\in\F_i$ with $s_j(X(T_2))=0$, property~(\ref{prop:state-unchanged}) implies that $s_j(X(T_4))=0$. Moreover, by Eq.~\eqref{eq:conv-T2T4-basic} and the definition of $X^+(t)$, we have $s_j(X^+(T_4))=s_j(X^+(T_2))$, and hence $u_{ji}(X^+(T_4))=u_{ji}(X^+(T_2))$. Therefore, there are at least $\Delta u+1$ friends of $i$ such that $s_j(X(T_2))>0$, $s_j(X^+(T_2))<0$, and $s_j(X^+(T_4))\ge 0$. Denote such friends of $i$ by $r_1,\dots, r_m$, with $m\ge \Delta u +1$. In particular, since $s_i(X(T_2))=0$, $i$ itself is not in the set $\{r_1,\dots, r_m\}$. For each $k\in\{1,\dots,m\}$, there must exist $t_k\in \{T_2+1,\dots, T_4\}$ such that 

\begin{align}
        \notag s_{r_k}(X^+(t_k-1))  &= s_{r_k}(X(t_k-1)) \\
        \label{eq:conv-tk} &\quad - x_{i r_k}(t_k-1) + x_{ir_k}^+(T_4)<0,\\
        \notag s_{r_k}(X^+(t_k))  &= s_{r_k}(X(t_k)) - x_{ir_k}(t_k) + x_{ir_k}^+(T_4)\ge 0.
\end{align}

Let $t^*=\max\{t_1,\dots, t_m\}$ and $\V^* = \{r_k\,|\, k\in \{1,\dots, m\},\,\, t_k = t^*\}$. By definition, $\V^*$ is non-empty and is a subset of $\{r_1,\dots, r_m\}$. Meanwhile, it follows that $u_i(X^+(t^*))\ge u_i(X^+(T_4))> u_i(X(T_4))= u_i(X(t^*))$.

Fix any $l\in \V^*$. In the two scenarios below, we have $s_l\big(X^+(t^*-1)\big)<0$ and $s_l\big(X^+(t^*)\big)\ge 0$. We next show that the latter inequality holds with equality, i.e., $s_l(X^+(t^*))=0$.

\emph{Scenario 1:} $i_{t^*}=l$. Since $x_{il}(t^*-1)=x_{il}(t^*)$, inequality~\eqref{eq:conv-tk} implies that $s_l(X(t^*-1))<s_l(X(t^*))$. According to Rule~3, $s_l(X(t^*))=s_l(X(t^*-1))+1$, and thus $s_l(X^+(t^*-1))=-1$ and $s_l(X^+(t^*))=0$.

\emph{Scenario 2:} $i_{t^*}\neq l$. Country $i_{t^*}$ cannot be a friend of $l$, since otherwise Rule~3 would imply $x_{i_{t^*}l}(t^*-1)=x_{i_{t^*}l}(t^*)$ and hence $s_l(X(t^*-1))=s_l(X(t^*))$, which contradicts the assumed state transition. Therefore, $i_{t^*}\in \A_l$. Since $x_{kl}(t^*-1)=x_{kl}(t^*)$ for any $k\neq i_{t^*}$, Rule~3 implies that $i_{t^*}$ retracts power from attacking $l$, resulting in $s_l(X(t^*-1))<s_l(X(t^*))$. Notice that each retraction performed in Phases~3 and~4 of
Algorithm~\ref{alg:trajectory} involves exactly one unit of
power. Therefore, $x_{i_{t^*}l}(t^*)=x_{i_{t^*}l}(t^*-1)-1$, implying that $s_l(X(t^*))=s_l(X(t^*-1))+1$. Therefore, we have $s_l(X^+(t^*-1))=-1$ and $s_l(X^+(t^*))=0$.

The strategy matrix obtained by directly applying the
$i$-th row of $X^+(T_4)$ at time $t^*$ need not satisfy
Rule~1. We therefore truncate this allocation according to
the upper bounds imposed by Rule~1. Define
\[
c_{ij}=
\begin{cases}
x_{ij}(t^*), & u_{ji}(X(t^*))=1,\\
x_{ij}(t^*)+|s_j(X(t^*))|, & u_{ji}(X(t^*))=0.
\end{cases}
\]
Set $\widehat x^+_{ij}= \min\{x^+_{ij}(T_4),c_{ij}\}$,
for all $j\neq i$, and allocate the remaining power to country $i$ itself. All other rows remain identical to those of $X(t^*)$. By construction, $u_i(\widehat X^+(t^*))\ge u_i(X^+(t^*))\ge u_i(X(t^*))+1$. Therefore, $\widehat X^+(t^*)\in\bfU_i(X(t^*))$.

Let $\C_i^0=\C_i^0(\widehat X^+(t^*),X(t^*))$, where the definition of $\C_i^0(\widehat X^+(t^*),X(t^*))$ is given by Eq.~\eqref{eq:def} in Appendix II. Notice that there exists $l\in\{r_1,\dots, r_m\}$ satisfying 
\begin{align*}
    s_l(X(t^*))\ge s_l(X(T_2))>0,\text{ and }s_l(\widehat X^+(t^*))=0.
\end{align*}
Therefore, $N_p(\C_i^0,\widehat X^+(t^*))-N_p(\C_i^0,X(t^*))\ge 1$. By Eq.~\eqref{eq:prenum} in Lemma~\ref{lem:precarious number}, 
\begin{align*}
    N_p&(\V,\widehat X^+(t^*))-N_p(\V,X(t^*)) \\
    &\ge N_p(\V\setminus\{i\},\widehat X^+(t^*))-N_p(\V\setminus\{i\},X(t^*))-1\\
    &\ge u_i(\widehat X^+(t^*))-u_i(X(t^*))+1-1 \ge 1.
\end{align*}
However, property~(\ref{prop:belong_S1}) implies that $X(t^*)\in\bfS_1$. Since country $i$ is precarious and $\widehat X^+(t^*)\in\bfU_i(X(t^*))$, the strict increase in the number of precarious countries contradicts the definition of $\bfS_1$.

This contradicts the assumption that there exist $i$ and $X^+(T_4)\in \bfU_i(X(T_4))$ satisfying $u_i(X^+(T_4))\ge u_i(X(T_4))+1$. That is, the matrix $X(T_4)$ produced by Algorithm~\ref{alg:trajectory} in finite time is an equilibrium of the COPS dynamics. Combining this conclusion with Lemmas~\ref{lem:Markev} and~\ref{lem:no-precarious-initially}, we conclude that the COPS dynamics almost surely reaches an equilibrium in finite time. This concludes the proof.\hfill$\blacksquare$

\section{Simulation results during World War II}\label{apx:WW2}
This section provides the detailed data underlying the World War II case study in Section~IV-A. Throughout Appendices~VII and VIII, the units of analysis in the signed-network data are states as coded by the Correlates of War project, while country names in the postwar economic tables follow the nomenclature of the Penn World Table. Historical or standardized names are retained where applicable. For each year from 1939 to 1945, we first report the signed relationships among the eight selected countries, where $1$, $-1$, and $0$ represent friendly, antagonistic, and absent relationships, respectively. We then report each country's historical state and the survival likelihood computed from the simulations, thereby allowing a direct comparison between the model predictions and the corresponding historical observations. In each table associated with predictor year $t$, the survival likelihood is computed from year-$t$ data, whereas the historical state is determined according to the corresponding empirical outcome. For a given predictor year, the historical state equals $1$ if no war occurs within the country's territory in the following year, and equals $0$ otherwise.

\input{relation_tables/1939}
\input{relation_tables/1940.tex}
\input{relation_tables/1941.tex}
\input{relation_tables/1942.tex}
\input{relation_tables/1943.tex}
\input{relation_tables/1944.tex}
\input{relation_tables/1945.tex}
\FloatBarrier
\clearpage

\input{latex_tables/table_1939.tex}
\input{latex_tables/table_1940.tex}
\input{latex_tables/table_1941.tex}
\input{latex_tables/table_1942.tex}
\clearpage
\input{latex_tables/table_1943.tex}
\input{latex_tables/table_1944.tex}
\input{latex_tables/table_1945.tex}
\FloatBarrier

\section{Postwar simulation results}\label{apx:PW}
This section provides supplementary results for the postwar empirical study in Section~IV-A. We first present the signed relationships among the eight selected countries in 1955, 1967, 1979, and 1991. For each of these four selected years, we then report the simulated survival likelihood of each country together with an indicator of its subsequent empirical power growth. The growth indicator equals $1$ if the GDP-based proxy for national power increases in the subsequent year and equals $0$ otherwise. Finally, we report the annual prediction accuracy from 1955 to 2002, providing a more detailed assessment of the relationship between countries' survival likelihoods and subsequent changes in national power.

\input{relation_tables/1955.tex}
\input{relation_tables/1967.tex}
\input{relation_tables/1979.tex}
\input{relation_tables/1991.tex}

\input{growth_rate_tex_tables/table_growth_1955.tex}
\input{growth_rate_tex_tables/table_growth_1967.tex}
\input{growth_rate_tex_tables/table_growth_1979.tex}
\clearpage
\input{growth_rate_tex_tables/table_growth_1991.tex}

\FloatBarrier
\input{accuracy_1955_2002.tex}

\section*{References}
\bibliographystyle{IEEEtran} 
\bibliography{ref}

\end{document}

%% file: relation_tables/1939.tex
\begin{table*}[t]
  \centering
  \caption{Relations among selected countries in 1939}
  \label{tab:major-country-relations-1939}
  \renewcommand{\arraystretch}{1.2}
  \resizebox{\textwidth}{!}{%
  \begin{tabular}{ccccccccc}
    \toprule
    \textbf{Country} & \textbf{China} & \textbf{United States} & \textbf{Japan} & \textbf{United Kingdom} & \textbf{France} & \textbf{Germany} & \textbf{USSR} & \textbf{Italy} \\
    \midrule
    \textbf{China} & 1 & 0 & -1 & 0 & 0 & 0 & 1 & 0 \\
    \textbf{United States} & 0 & 1 & 0 & -1 & 0 & -1 & 0 & 0 \\
    \textbf{Japan} & -1 & 0 & 1 & -1 & -1 & 1 & -1 & 1 \\
    \textbf{United Kingdom} & 0 & -1 & -1 & 1 & 1 & -1 & -1 & -1 \\
    \textbf{France} & 0 & 0 & -1 & 1 & 1 & -1 & -1 & -1 \\
    \textbf{Germany} & 0 & -1 & 1 & -1 & -1 & 1 & 1 & -1 \\
    \textbf{USSR} & 1 & 0 & -1 & -1 & -1 & 1 & 1 & 1 \\
    \textbf{Italy} & 0 & 0 & 1 & -1 & -1 & -1 & 1 & 1 \\
    \bottomrule
  \end{tabular}%
  }
\end{table*}

%% file: relation_tables/1940.tex
\begin{table*}[t]
  \centering
  \caption{Relations among selected countries in 1940}
  \label{tab:major-country-relations-1940}
  \renewcommand{\arraystretch}{1.2}
  \resizebox{\textwidth}{!}{%
  \begin{tabular}{ccccccccc}
    \toprule
    \textbf{Country} & \textbf{China} & \textbf{United States} & \textbf{Japan} & \textbf{United Kingdom} & \textbf{France} & \textbf{Germany} & \textbf{USSR} & \textbf{Italy} \\
    \midrule
    \textbf{China} & 1 & 0 & -1 & 0 & 0 & 0 & 1 & 0 \\
    \textbf{United States} & 0 & 1 & -1 & -1 & 0 & 0 & 0 & 0 \\
    \textbf{Japan} & -1 & -1 & 1 & -1 & -1 & 1 & 0 & 1 \\
    \textbf{United Kingdom} & 0 & -1 & -1 & 1 & -1 & -1 & -1 & -1 \\
    \textbf{France} & 0 & 0 & -1 & -1 & 1 & -1 & 0 & -1 \\
    \textbf{Germany} & 0 & 0 & 1 & -1 & -1 & 1 & -1 & -1 \\
    \textbf{USSR} & 1 & 0 & 0 & -1 & 0 & -1 & 1 & 1 \\
    \textbf{Italy} & 0 & 0 & 1 & -1 & -1 & -1 & 1 & 1 \\
    \bottomrule
  \end{tabular}%
  }
\end{table*}

%% file: relation_tables/1941.tex
\begin{table*}[t]
  \centering
  \caption{Relations among selected countries in 1941}
  \label{tab:major-country-relations-1941}
  \renewcommand{\arraystretch}{1.2}
  \resizebox{\textwidth}{!}{%
  \begin{tabular}{ccccccccc}
    \toprule
    \textbf{Country} & \textbf{China} & \textbf{United States} & \textbf{Japan} & \textbf{United Kingdom} & \textbf{France} & \textbf{Germany} & \textbf{USSR} & \textbf{Italy} \\
    \midrule
    \textbf{China} & 1 & 0 & -1 & 0 & 0 & 0 & 1 & 0 \\
    \textbf{United States} & 0 & 1 & -1 & 0 & 0 & -1 & 0 & -1 \\
    \textbf{Japan} & -1 & -1 & 1 & -1 & 0 & 1 & 1 & 1 \\
    \textbf{United Kingdom} & 0 & 0 & -1 & 1 & -1 & -1 & 1 & -1 \\
    \textbf{France} & 0 & 0 & 0 & -1 & 1 & 0 & 0 & 0 \\
    \textbf{Germany} & 0 & -1 & 1 & -1 & 0 & 1 & -1 & 1 \\
    \textbf{USSR} & 1 & 0 & 1 & 1 & 0 & -1 & 1 & 1 \\
    \textbf{Italy} & 0 & -1 & 1 & -1 & 0 & 1 & 1 & 1 \\
    \bottomrule
  \end{tabular}%
  }
\end{table*}

%% file: relation_tables/1942.tex
\begin{table*}[t]
  \centering
  \caption{Relations among selected countries in 1942}
  \label{tab:major-country-relations-1942}
  \renewcommand{\arraystretch}{1.2}
  \resizebox{\textwidth}{!}{%
  \begin{tabular}{ccccccccc}
    \toprule
    \textbf{Country} & \textbf{China} & \textbf{United States} & \textbf{Japan} & \textbf{United Kingdom} & \textbf{France} & \textbf{Germany} & \textbf{USSR} & \textbf{Italy} \\
    \midrule
    \textbf{China} & 1 & 0 & -1 & 0 & 0 & 0 & 1 & 0 \\
    \textbf{United States} & 0 & 1 & -1 & 0 & -1 & -1 & 0 & -1 \\
    \textbf{Japan} & -1 & -1 & 1 & -1 & 0 & 0 & 1 & 0 \\
    \textbf{United Kingdom} & 0 & 0 & -1 & 1 & -1 & -1 & 1 & -1 \\
    \textbf{France} & 0 & -1 & 0 & -1 & 1 & 0 & 0 & 0 \\
    \textbf{Germany} & 0 & -1 & 0 & -1 & 0 & 1 & -1 & 1 \\
    \textbf{USSR} & 1 & 0 & 1 & 1 & 0 & -1 & 1 & 0 \\
    \textbf{Italy} & 0 & -1 & 0 & -1 & 0 & 1 & 0 & 1 \\
    \bottomrule
  \end{tabular}%
  }
\end{table*}

%% file: relation_tables/1943.tex
\begin{table*}[t]
  \centering
  \caption{Relations among selected countries in 1943}
  \label{tab:major-country-relations-1943}
  \renewcommand{\arraystretch}{1.2}
  \resizebox{\textwidth}{!}{%
  \begin{tabular}{cccccccc}
    \toprule
    \textbf{Country} & \textbf{China} & \textbf{United States} & \textbf{Japan} & \textbf{United Kingdom} & \textbf{Germany} & \textbf{USSR} & \textbf{Italy} \\
    \midrule
    \textbf{China} & 1 & 0 & -1 & 0 & 0 & -1 & 0 \\
    \textbf{United States} & 0 & 1 & -1 & 0 & -1 & 0 & -1 \\
    \textbf{Japan} & -1 & -1 & 1 & -1 & 0 & 1 & 0 \\
    \textbf{United Kingdom} & 0 & 0 & -1 & 1 & -1 & 1 & -1 \\
    \textbf{Germany} & 0 & -1 & 0 & -1 & 1 & -1 & -1 \\
    \textbf{USSR} & -1 & 0 & 1 & 1 & -1 & 1 & 0 \\
    \textbf{Italy} & 0 & -1 & 0 & -1 & -1 & 0 & 1 \\
    \bottomrule
  \end{tabular}%
  }
\end{table*}

%% file: relation_tables/1944.tex
\begin{table*}[t]
  \centering
  \caption{Relations among selected countries in 1944}
  \label{tab:major-country-relations-1944}
  \renewcommand{\arraystretch}{1.2}
  \resizebox{\textwidth}{!}{%
  \begin{tabular}{ccccccccc}
    \toprule
    \textbf{Country} & \textbf{China} & \textbf{United States} & \textbf{Japan} & \textbf{United Kingdom} & \textbf{France} & \textbf{Germany} & \textbf{USSR} & \textbf{Italy} \\
    \midrule
    \textbf{China} & 1 & 0 & -1 & 0 & 0 & 0 & -1 & 0 \\
    \textbf{United States} & 0 & 1 & -1 & 0 & 0 & -1 & 0 & 0 \\
    \textbf{Japan} & -1 & -1 & 1 & -1 & -1 & 0 & 1 & 0 \\
    \textbf{United Kingdom} & 0 & 0 & -1 & 1 & 0 & -1 & 1 & 0 \\
    \textbf{France} & 0 & 0 & -1 & 0 & 1 & -1 & 1 & 0 \\
    \textbf{Germany} & 0 & -1 & 0 & -1 & -1 & 1 & -1 & -1 \\
    \textbf{USSR} & -1 & 0 & 1 & 1 & 1 & -1 & 1 & 0 \\
    \textbf{Italy} & 0 & 0 & 0 & 0 & 0 & -1 & 0 & 1 \\
    \bottomrule
  \end{tabular}%
  }
\end{table*}

%% file: relation_tables/1945.tex
\begin{table*}[t]
  \centering
  \caption{Relations among selected countries in 1945}
  \label{tab:major-country-relations-1945}
  \renewcommand{\arraystretch}{1.2}
  \resizebox{\textwidth}{!}{%
  \begin{tabular}{ccccccccc}
    \toprule
    \textbf{Country} & \textbf{China} & \textbf{United States} & \textbf{Japan} & \textbf{United Kingdom} & \textbf{France} & \textbf{Germany} & \textbf{USSR} & \textbf{Italy} \\
    \midrule
    \textbf{China} & 1 & 0 & -1 & 0 & 0 & 0 & -1 & 0 \\
    \textbf{United States} & 0 & 1 & -1 & 0 & 0 & -1 & 0 & 0 \\
    \textbf{Japan} & -1 & -1 & 1 & -1 & 0 & 0 & -1 & 0 \\
    \textbf{United Kingdom} & 0 & 0 & -1 & 1 & 0 & -1 & 1 & 0 \\
    \textbf{France} & 0 & 0 & 0 & 0 & 1 & -1 & 1 & 0 \\
    \textbf{Germany} & 0 & -1 & 0 & -1 & -1 & 1 & -1 & -1 \\
    \textbf{USSR} & -1 & 0 & -1 & 1 & 1 & -1 & 1 & 0 \\
    \textbf{Italy} & 0 & 0 & 0 & 0 & 0 & -1 & 0 & 1 \\
    \bottomrule
  \end{tabular}%
  }
\end{table*}

%% file: latex_tables/table_1939.tex
\begin{table}[!t]
\centering
\caption{Survival likelihoods and historical states in 1939.}
\label{tab:ww2_1939}
\scriptsize
\setlength{\tabcolsep}{2.5pt}
\renewcommand{\arraystretch}{0.95}
\begin{adjustbox}{max width=\columnwidth}
\begin{tabular}{@{}lcc@{\hspace{0.8em}}lcc@{}}
\toprule
Country & \shortstack{Survival\\Likelihood} & \shortstack{Historical\\State} & Country & \shortstack{Survival\\Likelihood} & \shortstack{Historical\\State} \\
\midrule
United States & 0.98 & 1 & Germany & 1 & 1 \\
Canada & 0.13 & 1 & Poland & 0.845 & 0 \\
Cuba & 1 & 1 & Hungary & 1 & 1 \\
Haiti & 1 & 1 & Italy & 0.98 & 1 \\
Dominican Republic & 1 & 1 & Albania & 0 & 0 \\
Mexico & 1 & 1 & Yugoslavia & 1 & 1 \\
Guatemala & 1 & 1 & Greece & 1 & 1 \\
Honduras & 1 & 1 & Bulgaria & 1 & 1 \\
El Salvador & 1 & 1 & Romania & 1 & 1 \\
Nicaragua & 1 & 1 & USSR & 1 & 1 \\
Costa Rica & 1 & 1 & Estonia & 0 & 1 \\
Panama & 1 & 1 & Latvia & 0 & 1 \\
Colombia & 1 & 1 & Lithuania & 0.885 & 1 \\
Venezuela & 1 & 1 & Finland & 0.005 & 0 \\
Ecuador & 1 & 1 & Sweden & 0.075 & 1 \\
Peru & 1 & 1 & Norway & 0.005 & 1 \\
Brazil & 1 & 1 & Denmark & 1 & 1 \\
Bolivia & 1 & 1 & Liberia & 1 & 1 \\
Paraguay & 1 & 1 & South Africa & 0.03 & 1 \\
Chile & 1 & 1 & Iran & 1 & 1 \\
Argentina & 0.945 & 1 & Turkey & 1 & 1 \\
Uruguay & 0.63 & 1 & Iraq & 1 & 1 \\
United Kingdom & 0.885 & 1 & Egypt & 1 & 1 \\
Ireland & 1 & 1 & Saudi Arabia & 1 & 1 \\
Netherlands & 0.005 & 1 & Afghanistan & 0.025 & 1 \\
Belgium & 0.155 & 1 & China & 1 & 1 \\
Luxembourg & 0.03 & 1 & Japan & 0 & 1 \\
France & 0.42 & 1 & Nepal & 1 & 1 \\
Switzerland & 0.02 & 1 & Thailand & 1 & 1 \\
Spain & 0.95 & 0 & Australia & 0.08 & 1 \\
Portugal & 0.935 & 1 & New Zealand & 0 & 1 \\
\bottomrule
\end{tabular}
\end{adjustbox}
\end{table}

%% file: latex_tables/table_1940.tex
\begin{table}[!t]
\centering
\caption{Survival likelihoods and historical states in 1940.}
\label{tab:ww2_1940}
\scriptsize
\setlength{\tabcolsep}{2.5pt}
\renewcommand{\arraystretch}{0.95}
\begin{adjustbox}{max width=\columnwidth}
\begin{tabular}{@{}lcc@{\hspace{0.8em}}lcc@{}}
\toprule
Country & \shortstack{Survival\\Likelihood} & \shortstack{Historical\\State} & Country & \shortstack{Survival\\Likelihood} & \shortstack{Historical\\State} \\
\midrule
United States & 0.99 & 1 & Portugal & 0.8 & 1 \\
Canada & 0.625 & 1 & Germany & 1 & 1 \\
Cuba & 1 & 1 & Hungary & 0.825 & 1 \\
Haiti & 1 & 1 & Italy & 0.44 & 0 \\
Dominican Republic & 1 & 1 & Yugoslavia & 0.525 & 1 \\
Mexico & 1 & 1 & Greece & 0.82 & 0 \\
Guatemala & 1 & 1 & Bulgaria & 0.135 & 1 \\
Honduras & 1 & 1 & Romania & 0.28 & 1 \\
El Salvador & 1 & 1 & USSR & 1 & 1 \\
Nicaragua & 1 & 1 & Estonia & 0.115 & 0 \\
Costa Rica & 1 & 1 & Latvia & 0 & 0 \\
Panama & 0.955 & 1 & Lithuania & 0.01 & 0 \\
Colombia & 1 & 1 & Finland & 0 & 0 \\
Venezuela & 1 & 1 & Sweden & 0.02 & 1 \\
Ecuador & 1 & 1 & Norway & 0.005 & 0 \\
Peru & 1 & 1 & Denmark & 0.005 & 0 \\
Brazil & 0.94 & 1 & Liberia & 1 & 1 \\
Bolivia & 1 & 1 & South Africa & 0.075 & 1 \\
Paraguay & 1 & 1 & Iran & 0.99 & 1 \\
Chile & 1 & 1 & Turkey & 0.965 & 1 \\
Argentina & 0.925 & 1 & Iraq & 1 & 1 \\
Uruguay & 0.96 & 1 & Egypt & 0.995 & 1 \\
United Kingdom & 0.8 & 1 & Saudi Arabia & 1 & 1 \\
Ireland & 0.02 & 1 & Afghanistan & 0.06 & 1 \\
Netherlands & 0.145 & 0 & China & 1 & 1 \\
Belgium & 0.31 & 0 & Japan & 0 & 1 \\
Luxembourg & 0.06 & 0 & Nepal & 1 & 1 \\
France & 1 & 0 & Thailand & 0.75 & 1 \\
Switzerland & 0.045 & 1 & Australia & 0.195 & 1 \\
Spain & 0.15 & 1 & New Zealand & 0 & 1 \\
\bottomrule
\end{tabular}
\end{adjustbox}
\end{table}

%% file: latex_tables/table_1941.tex
\begin{table}[!t]
\centering
\caption{Survival likelihoods and historical states in 1941.}
\label{tab:ww2_1941}
\scriptsize
\setlength{\tabcolsep}{2.5pt}
\renewcommand{\arraystretch}{0.95}
\begin{adjustbox}{max width=\columnwidth}
\begin{tabular}{@{}lcc@{\hspace{0.8em}}lcc@{}}
\toprule
Country & \shortstack{Survival\\Likelihood} & \shortstack{Historical\\State} & Country & \shortstack{Survival\\Likelihood} & \shortstack{Historical\\State} \\
\midrule
United States & 1 & 1 & Portugal & 0.825 & 1 \\
Canada & 1 & 1 & Germany & 1 & 1 \\
Cuba & 0.51 & 1 & Hungary & 0.005 & 1 \\
Haiti & 0.44 & 1 & Italy & 0.925 & 0 \\
Dominican Republic & 0.57 & 1 & Yugoslavia & 0.92 & 0 \\
Mexico & 1 & 1 & Greece & 0.035 & 0 \\
Guatemala & 0.51 & 1 & Bulgaria & 0.005 & 1 \\
Honduras & 0.51 & 1 & Romania & 0 & 1 \\
El Salvador & 0.53 & 1 & USSR & 1 & 1 \\
Nicaragua & 0.445 & 1 & Finland & 0 & 0 \\
Costa Rica & 0.46 & 1 & Sweden & 0.12 & 1 \\
Panama & 0.52 & 1 & Liberia & 1 & 1 \\
Colombia & 1 & 1 & Ethiopia & 1 & 0 \\
Venezuela & 1 & 1 & South Africa & 0.04 & 1 \\
Ecuador & 1 & 1 & Iran & 0.025 & 0 \\
Peru & 1 & 1 & Turkey & 0.895 & 1 \\
Brazil & 1 & 1 & Iraq & 0.075 & 0 \\
Bolivia & 1 & 1 & Egypt & 0.885 & 1 \\
Paraguay & 1 & 1 & Saudi Arabia & 1 & 1 \\
Chile & 1 & 1 & Afghanistan & 1 & 1 \\
Argentina & 1 & 1 & China & 1 & 1 \\
Uruguay & 1 & 1 & Japan & 0 & 1 \\
United Kingdom & 1 & 1 & Nepal & 1 & 1 \\
Ireland & 1 & 1 & Thailand & 0 & 0 \\
France & 0.07 & 0 & Australia & 0 & 1 \\
Switzerland & 1 & 1 & New Zealand & 0 & 1 \\
Spain & 0.075 & 1 &  &  &  \\
\bottomrule
\end{tabular}
\end{adjustbox}
\end{table}

%% file: latex_tables/table_1942.tex
\begin{table}[!t]
\centering
\caption{Survival likelihoods and historical states in 1942.}
\label{tab:ww2_1942}
\scriptsize
\setlength{\tabcolsep}{2.5pt}
\renewcommand{\arraystretch}{0.95}
\begin{adjustbox}{max width=\columnwidth}
\begin{tabular}{@{}lcc@{\hspace{0.8em}}lcc@{}}
\toprule
Country & \shortstack{Survival\\Likelihood} & \shortstack{Historical\\State} & Country & \shortstack{Survival\\Likelihood} & \shortstack{Historical\\State} \\
\midrule
United States & 1 & 1 & Spain & 0 & 1 \\
Canada & 0.135 & 1 & Portugal & 1 & 1 \\
Cuba & 1 & 1 & Germany & 1 & 1 \\
Haiti & 0.9675 & 1 & Hungary & 0 & 1 \\
Dominican Republic & 1 & 1 & Italy & 0.2 & 0 \\
Mexico & 0.1275 & 1 & Bulgaria & 0 & 1 \\
Guatemala & 1 & 1 & Romania & 0 & 1 \\
Honduras & 0.7025 & 1 & USSR & 1 & 1 \\
El Salvador & 1 & 1 & Finland & 0 & 0 \\
Nicaragua & 1 & 1 & Sweden & 0.03 & 1 \\
Costa Rica & 1 & 1 & Liberia & 1 & 1 \\
Panama & 0.7525 & 1 & Ethiopia & 0 & 0 \\
Colombia & 1 & 1 & South Africa & 0 & 1 \\
Venezuela & 0.71 & 1 & Iran & 1 & 1 \\
Ecuador & 1 & 1 & Turkey & 1 & 1 \\
Peru & 1 & 1 & Iraq & 1 & 1 \\
Brazil & 0.49 & 1 & Egypt & 0 & 1 \\
Bolivia & 1 & 1 & Saudi Arabia & 1 & 1 \\
Paraguay & 1 & 1 & Afghanistan & 1 & 1 \\
Chile & 0.73 & 1 & China & 1 & 1 \\
Argentina & 0.7 & 1 & Japan & 0.0025 & 0 \\
Uruguay & 0.65 & 1 & Nepal & 1 & 1 \\
United Kingdom & 1 & 1 & Thailand & 0 & 0 \\
Ireland & 1 & 1 & Australia & 0.0125 & 1 \\
France & 0.01 & 0 & New Zealand & 0 & 1 \\
Switzerland & 1 & 1 &  &  &  \\
\bottomrule
\end{tabular}
\end{adjustbox}
\end{table}

%% file: latex_tables/table_1943.tex
\begin{table}[!t]
\centering
\caption{Survival likelihoods and historical states in 1943.}
\label{tab:ww2_1943}
\scriptsize
\setlength{\tabcolsep}{2.5pt}
\renewcommand{\arraystretch}{0.95}
\begin{adjustbox}{max width=\columnwidth}
\begin{tabular}{@{}lcc@{\hspace{0.8em}}lcc@{}}
\toprule
Country & \shortstack{Survival\\Likelihood} & \shortstack{Historical\\State} & Country & \shortstack{Survival\\Likelihood} & \shortstack{Historical\\State} \\
\midrule
United States & 1 & 1 & Spain & 0.065 & 1 \\
Canada & 0.215 & 1 & Portugal & 1 & 1 \\
Cuba & 1 & 1 & Germany & 1 & 1 \\
Haiti & 1 & 1 & Hungary & 0.06 & 1 \\
Dominican Republic & 1 & 1 & Italy & 0.01 & 0 \\
Mexico & 1 & 1 & Bulgaria & 0 & 1 \\
Guatemala & 1 & 1 & Romania & 0 & 0 \\
Honduras & 1 & 1 & USSR & 0.27 & 1 \\
El Salvador & 1 & 1 & Finland & 0.01 & 0 \\
Nicaragua & 1 & 1 & Sweden & 0.02 & 1 \\
Costa Rica & 1 & 1 & Liberia & 1 & 1 \\
Panama & 0.945 & 1 & Ethiopia & 1 & 0 \\
Colombia & 0.96 & 1 & South Africa & 0 & 1 \\
Venezuela & 1 & 1 & Iran & 0.94 & 1 \\
Ecuador & 1 & 1 & Turkey & 1 & 1 \\
Peru & 1 & 1 & Iraq & 0.275 & 1 \\
Brazil & 0.93 & 1 & Egypt & 0.85 & 1 \\
Bolivia & 0.85 & 1 & Saudi Arabia & 0.005 & 1 \\
Paraguay & 1 & 1 & Afghanistan & 1 & 1 \\
Chile & 1 & 1 & China & 1 & 1 \\
Argentina & 1 & 1 & Japan & 0.12 & 0 \\
Uruguay & 1 & 1 & Nepal & 1 & 1 \\
United Kingdom & 1 & 1 & Thailand & 0.005 & 0 \\
Ireland & 1 & 1 & Australia & 0.06 & 1 \\
Switzerland & 0.05 & 1 & New Zealand & 0 & 1 \\
\bottomrule
\end{tabular}
\end{adjustbox}
\end{table}

%% file: latex_tables/table_1944.tex
\begin{table}[!t]
\centering
\caption{Survival likelihoods and historical states in 1944.}
\label{tab:ww2_1944}
\scriptsize
\setlength{\tabcolsep}{2.5pt}
\renewcommand{\arraystretch}{0.95}
\begin{adjustbox}{max width=\columnwidth}
\begin{tabular}{@{}lcc@{\hspace{0.8em}}lcc@{}}
\toprule
Country & \shortstack{Survival\\Likelihood} & \shortstack{Historical\\State} & Country & \shortstack{Survival\\Likelihood} & \shortstack{Historical\\State} \\
\midrule
United States & 1 & 1 & Portugal & 1 & 1 \\
Canada & 0 & 1 & Germany & 1 & 0 \\
Cuba & 1 & 1 & Hungary & 0 & 0 \\
Haiti & 1 & 1 & Italy & 0 & 1 \\
Dominican Republic & 1 & 1 & Albania & 1 & 1 \\
Mexico & 1 & 1 & Yugoslavia & 1 & 0 \\
Guatemala & 1 & 1 & Greece & 1 & 1 \\
Honduras & 1 & 1 & Bulgaria & 0 & 0 \\
El Salvador & 1 & 1 & Romania & 0 & 0 \\
Nicaragua & 1 & 1 & USSR & 0.2 & 1 \\
Costa Rica & 1 & 1 & Finland & 0 & 0 \\
Panama & 1 & 1 & Sweden & 0 & 1 \\
Colombia & 1 & 1 & Iceland & 1 & 1 \\
Venezuela & 1 & 1 & Liberia & 0 & 1 \\
Ecuador & 1 & 1 & Ethiopia & 1 & 1 \\
Peru & 1 & 1 & South Africa & 0 & 1 \\
Brazil & 1 & 1 & Iran & 1 & 1 \\
Bolivia & 1 & 1 & Turkey & 1 & 1 \\
Paraguay & 1 & 1 & Iraq & 1 & 1 \\
Chile & 1 & 1 & Egypt & 1 & 1 \\
Argentina & 1 & 1 & Saudi Arabia & 1 & 1 \\
Uruguay & 1 & 1 & Afghanistan & 1 & 1 \\
United Kingdom & 1 & 1 & China & 1 & 1 \\
Ireland & 1 & 1 & Japan & 0 & 0 \\
Luxembourg & 1 & 0 & Nepal & 1 & 1 \\
France & 0.99 & 0 & Thailand & 0 & 0 \\
Switzerland & 1 & 1 & Australia & 0 & 1 \\
Spain & 0 & 1 & New Zealand & 0 & 1 \\
\bottomrule
\end{tabular}
\end{adjustbox}
\end{table}

%% file: latex_tables/table_1945.tex
\begin{table}[H]
\centering
\caption{Survival likelihoods and historical states in 1945.}
\label{tab:ww2_1945}
\scriptsize
\setlength{\tabcolsep}{2.5pt}
\renewcommand{\arraystretch}{0.95}
\begin{adjustbox}{max width=\columnwidth}
\begin{tabular}{@{}lcc@{\hspace{0.8em}}lcc@{}}
\toprule
Country & \shortstack{Survival\\Likelihood} & \shortstack{Historical\\State} & Country & \shortstack{Survival\\Likelihood} & \shortstack{Historical\\State} \\
\midrule
United States & 1 & 1 & Germany & 0 & 0 \\
Canada & 1 & 1 & Poland & 1 & 0 \\
Cuba & 1 & 1 & Hungary & 0 & 0 \\
Haiti & 1 & 1 & Italy & 1 & 1 \\
Dominican Republic & 1 & 1 & Albania & 1 & 1 \\
Mexico & 1 & 1 & Yugoslavia & 1 & 1 \\
Guatemala & 1 & 1 & Greece & 1 & 1 \\
Honduras & 1 & 1 & Bulgaria & 0 & 1 \\
El Salvador & 1 & 1 & Romania & 0 & 1 \\
Nicaragua & 1 & 1 & USSR & 1 & 1 \\
Costa Rica & 1 & 1 & Finland & 0 & 0 \\
Panama & 1 & 1 & Sweden & 1 & 1 \\
Colombia & 1 & 1 & Norway & 1 & 1 \\
Venezuela & 1 & 1 & Denmark & 1 & 1 \\
Ecuador & 1 & 1 & Iceland & 1 & 1 \\
Peru & 1 & 1 & Liberia & 1 & 1 \\
Brazil & 0.995 & 1 & Ethiopia & 1 & 1 \\
Bolivia & 1 & 1 & South Africa & 0.005 & 1 \\
Paraguay & 1 & 1 & Iran & 0.42 & 1 \\
Chile & 1 & 1 & Turkey & 0.305 & 1 \\
Argentina & 1 & 1 & Iraq & 1 & 1 \\
Uruguay & 1 & 1 & Egypt & 0.99 & 1 \\
United Kingdom & 1 & 1 & Saudi Arabia & 0.05 & 1 \\
Ireland & 1 & 1 & Afghanistan & 1 & 1 \\
Netherlands & 1 & 0 & China & 1 & 1 \\
Belgium & 1 & 0 & Japan & 0 & 0 \\
Luxembourg & 1 & 0 & Nepal & 1 & 1 \\
France & 1 & 0 & Thailand & 0 & 0 \\
Switzerland & 1 & 1 & Australia & 0.11 & 1 \\
Spain & 1 & 1 & New Zealand & 0.005 & 1 \\
Portugal & 1 & 1 &  &  &  \\
\bottomrule
\end{tabular}
\end{adjustbox}
\end{table}

%% file: relation_tables/1955.tex
\begin{table*}[t]
  \centering
  \caption{Relations among selected countries in 1955}
  \label{tab:major-country-relations-1955}
  \renewcommand{\arraystretch}{1.1}
  \resizebox{\textwidth}{!}{%
  \begin{tabular}{ccccccccc}
    \toprule
    \textbf{Country} & \textbf{China} & \textbf{United States} & \textbf{Japan} & \textbf{United Kingdom} & \textbf{France} & \textbf{USSR} & \textbf{Italy} & \textbf{Canada} \\
    \midrule
    \textbf{China} & 1 & -1 & 0 & 0 & 0 & 1 & 0 & 0 \\
    \textbf{United States} & -1 & 1 & 1 & 1 & 1 & 0 & 1 & 1 \\
    \textbf{Japan} & 0 & 1 & 1 & 0 & 0 & -1 & 0 & 0 \\
    \textbf{United Kingdom} & 0 & 1 & 0 & 1 & 1 & 1 & 1 & 1 \\
    \textbf{France} & 0 & 1 & 0 & 1 & 1 & 1 & 1 & 1 \\
    \textbf{USSR} & 1 & 0 & -1 & 1 & 1 & 1 & 0 & 0 \\
    \textbf{Italy} & 0 & 1 & 0 & 1 & 1 & 0 & 1 & 1 \\
    \textbf{Canada} & 0 & 1 & 0 & 1 & 1 & 0 & 1 & 1 \\
    \bottomrule
  \end{tabular}%
  }
\end{table*}

%% file: relation_tables/1967.tex
\begin{table*}[t]
  \centering
  \caption{Relations among selected countries in 1967}
  \label{tab:major-country-relations-1967}
  \renewcommand{\arraystretch}{1.1}
  \resizebox{\textwidth}{!}{%
  \begin{tabular}{ccccccccc}
    \toprule
    \textbf{Country} & \textbf{China} & \textbf{United States} & \textbf{Japan} & \textbf{United Kingdom} & \textbf{France} & \textbf{USSR} & \textbf{Italy} & \textbf{Canada} \\
    \midrule
    \textbf{China} & 1 & -1 & 0 & 0 & 0 & -1 & 0 & 0 \\
    \textbf{United States} & -1 & 1 & 1 & 1 & 1 & -1 & 1 & 1 \\
    \textbf{Japan} & 0 & 1 & 1 & 0 & 0 & -1 & 0 & 0 \\
    \textbf{United Kingdom} & 0 & 1 & 0 & 1 & 1 & -1 & 1 & 1 \\
    \textbf{France} & 0 & 1 & 0 & 1 & 1 & 0 & 1 & 1 \\
    \textbf{USSR} & -1 & -1 & -1 & -1 & 0 & 1 & 0 & 0 \\
    \textbf{Italy} & 0 & 1 & 0 & 1 & 1 & 0 & 1 & 1 \\
    \textbf{Canada} & 0 & 1 & 0 & 1 & 1 & 0 & 1 & 1 \\
    \bottomrule
  \end{tabular}%
  }
\end{table*}

%% file: relation_tables/1979.tex
\begin{table*}[t]
  \centering
  \caption{Relations among selected countries in 1979}
  \label{tab:major-country-relations-1979}
  \renewcommand{\arraystretch}{1.1}
  \resizebox{\textwidth}{!}{%
  \begin{tabular}{ccccccccc}
    \toprule
    \textbf{Country} & \textbf{China} & \textbf{United States} & \textbf{Japan} & \textbf{United Kingdom} & \textbf{France} & \textbf{USSR} & \textbf{Italy} & \textbf{Canada} \\
    \midrule
    \textbf{China} & 1 & 0 & 0 & 0 & 0 & -1 & 0 & 0 \\
    \textbf{United States} & 0 & 1 & 1 & 1 & 1 & -1 & 1 & -1 \\
    \textbf{Japan} & 0 & 1 & 1 & 0 & 0 & -1 & 0 & 0 \\
    \textbf{United Kingdom} & 0 & 1 & 0 & 1 & 1 & 0 & 1 & 1 \\
    \textbf{France} & 0 & 1 & 0 & 1 & 1 & 1 & 1 & 1 \\
    \textbf{USSR} & -1 & -1 & -1 & 0 & 1 & 1 & 0 & 1 \\
    \textbf{Italy} & 0 & 1 & 0 & 1 & 1 & 0 & 1 & 1 \\
    \textbf{Canada} & 0 & -1 & 0 & 1 & 1 & 1 & 1 & 1 \\
    \bottomrule
  \end{tabular}%
  }
\end{table*}

%% file: relation_tables/1991.tex
\begin{table*}[t]
  \centering
  \caption{Relations among selected countries in 1991}
  \label{tab:major-country-relations-1991}
  \renewcommand{\arraystretch}{1.1}
  \resizebox{\textwidth}{!}{%
  \begin{tabular}{ccccccccc}
    \toprule
    \textbf{Country} & \textbf{China} & \textbf{United States} & \textbf{Japan} & \textbf{United Kingdom} & \textbf{France} & \textbf{USSR} & \textbf{Italy} & \textbf{Canada} \\
    \midrule
    \textbf{China} & 1 & 0 & 0 & 0 & 0 & 0 & 0 & 0 \\
    \textbf{United States} & 0 & 1 & 1 & 1 & 1 & 0 & 1 & -1 \\
    \textbf{Japan} & 0 & 1 & 1 & 0 & 0 & 0 & 0 & 0 \\
    \textbf{United Kingdom} & 0 & 1 & 0 & 1 & 1 & 0 & 1 & 1 \\
    \textbf{France} & 0 & 1 & 0 & 1 & 1 & 1 & 1 & 1 \\
    \textbf{USSR} & 0 & 0 & 0 & 0 & 1 & 1 & 1 & 1 \\
    \textbf{Italy} & 0 & 1 & 0 & 1 & 1 & 1 & 1 & 1 \\
    \textbf{Canada} & 0 & -1 & 0 & 1 & 1 & 1 & 1 & 1 \\
    \bottomrule
  \end{tabular}%
  }
\end{table*}

%% file: growth_rate_tex_tables/table_growth_1955.tex
\begin{table}[!t]
\centering
\caption{Survival likelihoods and growth indicator in 1955.}
\label{tab:growth_1955}
\scriptsize
\setlength{\tabcolsep}{2.5pt}
\renewcommand{\arraystretch}{0.95}
\resizebox{\columnwidth}{!}{%
\begin{tabular}{@{}lcc@{\hspace{0.8em}}lcc@{}}
\toprule
Country & \shortstack{Survival\\Likelihood} & \shortstack{Growth\\Indicator} & Country & \shortstack{Survival\\Likelihood} & \shortstack{Growth\\Indicator} \\
\midrule
United States & 1 & 1 & Switzerland & 1 & 1 \\
Canada & 1 & 1 & Spain & 1 & 1 \\
Dominican Republic & 1 & 1 & Portugal & 1 & 1 \\
Mexico & 1 & 1 & Austria & 1 & 1 \\
Guatemala & 1 & 1 & Italy & 1 & 1 \\
Honduras & 1 & 1 & Greece & 1 & 1 \\
El Salvador & 1 & 1 & Finland & 1 & 1 \\
Nicaragua & 1 & 1 & Sweden & 0 & 1 \\
Costa Rica & 1 & 1 & Norway & 1 & 1 \\
Panama & 1 & 1 & Denmark & 1 & 0 \\
Colombia & 1 & 1 & Iceland & 1 & 1 \\
Venezuela & 1 & 1 & Ethiopia & 1 & 1 \\
Ecuador & 0 & 1 & South Africa & 1 & 1 \\
Peru & 0 & 1 & Turkey & 1 & 1 \\
Brazil & 1 & 1 & Egypt & 1 & 0 \\
Paraguay & 1 & 1 & Jordan & 1 & 0 \\
Chile & 1 & 1 & Israel & 0 & 1 \\
Argentina & 1 & 1 & China & 1 & 1 \\
Uruguay & 1 & 1 & Japan & 1 & 1 \\
United Kingdom & 1 & 1 & India & 1 & 1 \\
Ireland & 1 & 1 & Pakistan & 1 & 1 \\
Netherlands & 1 & 1 & Thailand & 1 & 0 \\
Belgium & 1 & 1 & Philippines & 1 & 1 \\
Luxembourg & 1 & 1 & Australia & 1 & 1 \\
France & 1 & 1 & New Zealand & 1 & 1 \\
\bottomrule
\end{tabular}
}
\end{table}

%% file: growth_rate_tex_tables/table_growth_1967.tex
\begin{table}[!t]
\centering
\caption{Survival likelihoods and growth indicator in 1967.}
\label{tab:growth_1967}
\scriptsize
\setlength{\tabcolsep}{2.5pt}
\renewcommand{\arraystretch}{0.95}
\resizebox{\columnwidth}{!}{%
\begin{tabular}{@{}lcc@{\hspace{0.8em}}lcc@{}}
\toprule
Country & \shortstack{Survival\\Likelihood} & \shortstack{Growth\\Indicator} & Country & \shortstack{Survival\\Likelihood} & \shortstack{Growth\\Indicator} \\
\midrule
United States & 1 & 1 & Benin & 1 & 1 \\
Canada & 1 & 1 & Mauritania & 1 & 1 \\
Haiti & 1 & 0 & Niger & 1 & 1 \\
Dominican Republic & 1 & 1 & Guinea & 1 & 0 \\
Jamaica & 1 & 1 & Burkina Faso & 1 & 1 \\
Barbados & 1 & 1 & Liberia & 1 & 1 \\
Mexico & 1 & 1 & Sierra Leone & 1 & 0 \\
Guatemala & 1 & 1 & Ghana & 1 & 1 \\
Honduras & 1 & 1 & Togo & 1 & 1 \\
El Salvador & 1 & 1 & Cameroon & 1 & 0 \\
Nicaragua & 1 & 1 & Nigeria & 1 & 0 \\
Costa Rica & 1 & 1 & Gabon & 1 & 1 \\
Panama & 1 & 1 & Central African Republic & 1 & 1 \\
Colombia & 1 & 1 & Chad & 1 & 1 \\
Venezuela & 1 & 1 & Congo & 1 & 1 \\
Ecuador & 0 & 1 & Uganda & 1 & 1 \\
Peru & 1 & 1 & Kenya & 1 & 1 \\
Brazil & 1 & 1 & Burundi & 1 & 1 \\
Paraguay & 1 & 1 & Ethiopia & 1 & 1 \\
Chile & 1 & 1 & Zambia & 0 & 1 \\
Argentina & 1 & 1 & Zimbabwe & 1 & 1 \\
Uruguay & 1 & 0 & Malawi & 1 & 0 \\
United Kingdom & 1 & 1 & South Africa & 1 & 1 \\
Ireland & 1 & 1 & Lesotho & 1 & 1 \\
Netherlands & 1 & 1 & Botswana & 1 & 1 \\
Belgium & 1 & 1 & Madagascar & 1 & 1 \\
Luxembourg & 1 & 1 & Algeria & 1 & 1 \\
France & 1 & 1 & Tunisia & 1 & 0 \\
Switzerland & 1 & 1 & Iran & 1 & 1 \\
Spain & 1 & 1 & Turkey & 1 & 1 \\
Portugal & 1 & 1 & Egypt & 1 & 0 \\
Austria & 1 & 1 & Jordan & 1 & 1 \\
Italy & 1 & 1 & Israel & 0 & 1 \\
Malta & 1 & 1 & China & 1 & 1 \\
Greece & 1 & 1 & Japan & 1 & 1 \\
Cyprus & 0 & 1 & India & 1 & 1 \\
Romania & 1 & 1 & Pakistan & 1 & 1 \\
Finland & 1 & 1 & Myanmar & 1 & 0 \\
Sweden & 1 & 1 & Nepal & 1 & 0 \\
Norway & 1 & 1 & Thailand & 1 & 1 \\
Denmark & 1 & 1 & Malaysia & 1 & 1 \\
Iceland & 1 & 0 & Philippines & 1 & 1 \\
Gambia & 1 & 0 & Indonesia & 1 & 1 \\
Mali & 1 & 0 & Australia & 1 & 1 \\
Senegal & 1 & 0 & New Zealand & 1 & 0 \\
\bottomrule
\end{tabular}
}
\end{table}

%% file: growth_rate_tex_tables/table_growth_1979.tex
\begin{table}[!t]
\centering
\caption{Survival likelihoods and growth indicator in 1979.}
\label{tab:growth_1979}
\scriptsize
\setlength{\tabcolsep}{2.5pt}
\renewcommand{\arraystretch}{0.95}
\resizebox{\columnwidth}{!}{%
\begin{tabular}{@{}lcc@{\hspace{0.8em}}lcc@{}}
\toprule
Country & \shortstack{Survival\\Likelihood} & \shortstack{Growth\\Indicator} & Country & \shortstack{Survival\\Likelihood} & \shortstack{Growth\\Indicator} \\
\midrule
United States & 1 & 1 & Niger & 1 & 1 \\
Canada & 1 & 1 & Guinea & 1 & 0 \\
Bahamas & 1 & 1 & Burkina Faso & 1 & 1 \\
Haiti & 1 & 1 & Liberia & 1 & 1 \\
Dominican Republic & 1 & 1 & Sierra Leone & 1 & 1 \\
Jamaica & 1 & 0 & Ghana & 1 & 0 \\
Barbados & 1 & 1 & Togo & 1 & 1 \\
Dominica & 1 & 0 & Cameroon & 1 & 1 \\
Mexico & 1 & 1 & Nigeria & 1 & 0 \\
Guatemala & 1 & 1 & Gabon & 1 & 1 \\
Honduras & 1 & 1 & Central African Republic & 1 & 0 \\
El Salvador & 1 & 0 & Chad & 1 & 0 \\
Nicaragua & 1 & 0 & Congo & 1 & 1 \\
Costa Rica & 1 & 1 & Uganda & 1 & 0 \\
Panama & 1 & 1 & Kenya & 1 & 1 \\
Colombia & 1 & 1 & Burundi & 1 & 1 \\
Venezuela & 1 & 1 & Djibouti & 1 & 1 \\
Guyana & 1 & 0 & Ethiopia & 1 & 1 \\
Suriname & 1 & 0 & Angola & 1 & 1 \\
Ecuador & 1 & 1 & Mozambique & 1 & 1 \\
Peru & 0 & 1 & Zambia & 1 & 0 \\
Brazil & 1 & 1 & Zimbabwe & 1 & 1 \\
Paraguay & 1 & 1 & Malawi & 1 & 1 \\
Chile & 1 & 1 & South Africa & 1 & 1 \\
Argentina & 1 & 1 & Lesotho & 1 & 0 \\
Uruguay & 1 & 1 & Botswana & 0 & 1 \\
United Kingdom & 1 & 1 & Swaziland & 1 & 1 \\
Ireland & 1 & 1 & Madagascar & 1 & 1 \\
Netherlands & 1 & 1 & Algeria & 1 & 1 \\
Belgium & 1 & 1 & Tunisia & 1 & 1 \\
Luxembourg & 1 & 1 & Sudan & 1 & 0 \\
France & 1 & 1 & Iran & 0 & 0 \\
Switzerland & 1 & 1 & Turkey & 1 & 0 \\
Spain & 1 & 1 & Iraq & 1 & 1 \\
Portugal & 1 & 1 & Egypt & 1 & 1 \\
Poland & 1 & 1 & Lebanon & 1 & 1 \\
Austria & 1 & 1 & Jordan & 1 & 1 \\
Hungary & 1 & 1 & Israel & 0 & 1 \\
Italy & 1 & 1 & Saudi Arabia & 1 & 1 \\
Malta & 1 & 1 & Kuwait & 1 & 1 \\
Albania & 1 & 1 & Bahrain & 1 & 1 \\
Greece & 1 & 1 & Qatar & 1 & 1 \\
Cyprus & 1 & 1 & United Arab Emirates & 1 & 1 \\
Bulgaria & 1 & 1 & Oman & 1 & 1 \\
Romania & 1 & 1 & China & 1 & 1 \\
Finland & 1 & 1 & Japan & 1 & 1 \\
Sweden & 1 & 1 & India & 1 & 0 \\
Norway & 1 & 1 & Bhutan & 1 & 1 \\
Denmark & 1 & 1 & Pakistan & 1 & 1 \\
Iceland & 1 & 1 & Bangladesh & 1 & 1 \\
Cape Verde & 1 & 1 & Myanmar & 1 & 1 \\
Sao Tome and Principe & 1 & 1 & Nepal & 1 & 1 \\
Guinea-Bissau & 1 & 1 & Thailand & 1 & 1 \\
Equatorial Guinea & 1 & 0 & Cambodia & 0 & 0 \\
Gambia & 1 & 0 & Malaysia & 0 & 1 \\
Mali & 1 & 1 & Philippines & 1 & 1 \\
Senegal & 1 & 1 & Indonesia & 1 & 1 \\
Benin & 1 & 1 & Australia & 1 & 1 \\
Mauritania & 1 & 1 & New Zealand & 1 & 1 \\
\bottomrule
\end{tabular}
}
\end{table}

%% file: growth_rate_tex_tables/table_growth_1991.tex
\begin{table}[H]
\centering
\caption{Survival likelihoods and growth indicator in 1991.}
\label{tab:growth_1991}
\scriptsize
\setlength{\tabcolsep}{2.5pt}
\renewcommand{\arraystretch}{0.95}
\resizebox{\columnwidth}{!}{%
\begin{tabular}{@{}lcc@{\hspace{0.8em}}lcc@{}}
\toprule
Country & \shortstack{Survival\\Likelihood} & \shortstack{Growth\\Indicator} & Country & \shortstack{Survival\\Likelihood} & \shortstack{Growth\\Indicator} \\
\midrule
United States & 1 & 0 & Niger & 0.99 & 1 \\
Canada & 1 & 0 & Guinea & 1 & 1 \\
Bahamas & 1 & 0 & Burkina Faso & 1 & 1 \\
Haiti & 1 & 1 & Liberia & 1 & 0 \\
Dominican Republic & 1 & 1 & Sierra Leone & 1 & 0 \\
Jamaica & 1 & 1 & Ghana & 1 & 1 \\
Barbados & 1 & 0 & Togo & 1 & 1 \\
Dominica & 1 & 1 & Cameroon & 1 & 0 \\
Mexico & 1 & 1 & Nigeria & 1 & 1 \\
Belize & 1 & 1 & Gabon & 1 & 1 \\
Guatemala & 1 & 1 & Central African Republic & 1 & 1 \\
Honduras & 1 & 1 & Chad & 1 & 1 \\
El Salvador & 1 & 1 & Congo & 1 & 1 \\
Nicaragua & 1 & 0 & Uganda & 1 & 1 \\
Costa Rica & 1 & 1 & Kenya & 1 & 1 \\
Panama & 1 & 1 & Burundi & 1 & 1 \\
Colombia & 1 & 1 & Djibouti & 1 & 0 \\
Venezuela & 1 & 1 & Ethiopia & 1 & 0 \\
Guyana & 1 & 1 & Angola & 1 & 1 \\
Suriname & 1 & 1 & Mozambique & 1 & 1 \\
Ecuador & 1 & 1 & Zambia & 1 & 0 \\
Peru & 1 & 1 & Zimbabwe & 1 & 1 \\
Brazil & 1 & 1 & Malawi & 1 & 1 \\
Paraguay & 1 & 1 & South Africa & 1 & 0 \\
Chile & 1 & 1 & Lesotho & 1 & 1 \\
Argentina & 1 & 1 & Botswana & 1 & 1 \\
Uruguay & 1 & 1 & Swaziland & 1 & 1 \\
United Kingdom & 1 & 0 & Madagascar & 1 & 0 \\
Ireland & 1 & 1 & Algeria & 1 & 0 \\
Netherlands & 1 & 1 & Tunisia & 1 & 1 \\
Belgium & 1 & 1 & Sudan & 1 & 1 \\
Luxembourg & 1 & 1 & Iran & 0 & 1 \\
France & 1 & 1 & Turkey & 1 & 1 \\
Switzerland & 1 & 0 & Iraq & 0 & 0 \\
Spain & 1 & 1 & Egypt & 1 & 1 \\
Portugal & 1 & 1 & Lebanon & 1 & 1 \\
Poland & 1 & 0 & Jordan & 0 & 1 \\
Austria & 1 & 1 & Israel & 1 & 1 \\
Hungary & 1 & 0 & Saudi Arabia & 1 & 1 \\
Italy & 1 & 1 & Kuwait & 1 & 0 \\
Malta & 1 & 1 & Bahrain & 1 & 1 \\
Albania & 1 & 0 & Qatar & 1 & 0 \\
Greece & 1 & 1 & United Arab Emirates & 1 & 1 \\
Cyprus & 1 & 1 & Oman & 1 & 1 \\
Bulgaria & 1 & 0 & China & 1 & 1 \\
Romania & 1 & 0 & Japan & 1 & 1 \\
Finland & 1 & 0 & India & 1 & 1 \\
Sweden & 1 & 0 & Bhutan & 1 & 1 \\
Norway & 1 & 1 & Pakistan & 1 & 1 \\
Denmark & 1 & 1 & Bangladesh & 1 & 1 \\
Iceland & 1 & 0 & Myanmar & 1 & 1 \\
Cape Verde & 1 & 1 & Nepal & 1 & 1 \\
Sao Tome and Principe & 1 & 1 & Thailand & 1 & 1 \\
Guinea-Bissau & 1 & 1 & Cambodia & 1 & 1 \\
Equatorial Guinea & 1 & 1 & Malaysia & 1 & 1 \\
Gambia & 1 & 1 & Philippines & 1 & 0 \\
Mali & 1 & 1 & Indonesia & 1 & 1 \\
Senegal & 1 & 0 & Australia & 1 & 1 \\
Benin & 1 & 1 & New Zealand & 1 & 0 \\
Mauritania & 1 & 1 &  &  &  \\
\bottomrule
\end{tabular}
}
\end{table}

%% file: accuracy_1955_2002.tex
\par\medskip

\captionof{table}{Annual prediction accuracy from 1955 to 2002.}
\label{tab:annual_accuracy}

\begingroup
\centering
\scriptsize
\setlength{\tabcolsep}{2.5pt}
\renewcommand{\arraystretch}{1.15}

\noindent
\resizebox{\columnwidth}{!}{%
\begin{tabular}{|c|c|c|c|c|c|c|c|c|}
\hline
\textbf{Year}
& 1955 & 1956 & 1957 & 1958
& 1959 & 1960 & 1961 & 1962 \\
\hline
\textbf{Accuracy (\%)}
& 83.67 & 92.00 & 90.38 & 76.92
& 78.84 & 82.45 & 84.93 & 82.05 \\
\hline
\end{tabular}%
}
\par\vspace{1.2mm}

\noindent
\resizebox{\columnwidth}{!}{%
\begin{tabular}{|c|c|c|c|c|c|c|c|c|}
\hline
\textbf{Year}
& 1963 & 1964 & 1965 & 1966
& 1967 & 1968 & 1969 & 1970 \\
\hline
\textbf{Accuracy (\%)}
& 87.50 & 87.95 & 88.37 & 83.14
& 77.52 & 83.33 & 86.66 & 91.11 \\
\hline
\end{tabular}%
}
\par\vspace{1.2mm}

\noindent
\resizebox{\columnwidth}{!}{%
\begin{tabular}{|c|c|c|c|c|c|c|c|c|}
\hline
\textbf{Year}
& 1971 & 1972 & 1973 & 1974
& 1975 & 1976 & 1977 & 1978 \\
\hline
\textbf{Accuracy (\%)}
& 90.74 & 80.55 & 82.56 & 82.72
& 66.95 & 81.73 & 81.03 & 76.92 \\
\hline
\end{tabular}%
}
\par\vspace{1.2mm}

\noindent
\resizebox{\columnwidth}{!}{%
\begin{tabular}{|c|c|c|c|c|c|c|c|c|}
\hline
\textbf{Year}
& 1979 & 1980 & 1981 & 1982
& 1983 & 1984 & 1985 & 1986 \\
\hline
\textbf{Accuracy (\%)}
& 80.34 & 75.21 & 67.79 & 64.40
& 58.47 & 78.81 & 76.27 & 77.96 \\
\hline
\end{tabular}%
}
\par\vspace{1.2mm}

\noindent
\resizebox{\columnwidth}{!}{%
\begin{tabular}{|c|c|c|c|c|c|c|c|c|}
\hline
\textbf{Year}
& 1987 & 1988 & 1989 & 1990
& 1991 & 1992 & 1993 & 1994 \\
\hline
\textbf{Accuracy (\%)}
& 77.11 & 80.50 & 79.66 & 76.27
& 74.57 & 77.11 & 73.72 & 85.59 \\
\hline
\end{tabular}%
}
\par\vspace{1.2mm}

\noindent
\resizebox{\columnwidth}{!}{%
\begin{tabular}{|c|c|c|c|c|c|c|c|c|}
\hline
\textbf{Year}
& 1995 & 1996 & 1997 & 1998
& 1999 & 2000 & 2001 & 2002 \\
\hline
\textbf{Accuracy (\%)}
& 88.13 & 88.98 & 87.28 & 81.35
& 77.96 & 82.20 & 83.89 & 84.74 \\
\hline
\end{tabular}%
}
\par

\endgroup

\medskip